\documentstyle[aps,prb,twocolumn,epsfig,eqsecnum,floats]{revtex}

\begin{document}
\bibliographystyle{../prsty}

\title{Interacting electrons with spin in a one--dimensional dirty wire
connected to leads} 
\author{In\`es Safi} \address{Service de Physique de
l'\'Etat Condens\'e, Centre d'\'Etudes de Saclay\\ 91191 Gif--sur--Yvette,
France\\} 

\author{H. J. Schulz} \address{Laboratoire de Physique des
Solides, Universit\'e Paris--Sud, 91405 Orsay, France}

\maketitle

\begin{abstract}
We investigate a one--dimensional wire of interacting electrons connected to
one--dimensional noninteracting leads in the absence and in the presence of
a backscattering potential. The ballistic wire separates the charge and spin
parts of an incident electron even in the noninteracting leads. The Fourier
transform of nonlocal correlation functions are computed for $T\gg
\omega$. In particular, this allows us to study the proximity effect,
related to the Andreev reflection. A new type of proximity effect emerges
when the wire has normally a tendency towards Wigner crystal formation. The
latter is suppressed by the leads below a space--dependent crossover
temperature; it gets dominated everywhere by the $2k_F$ CDW at $T<L^{\frac 3
2 (K-1) }$ for short range interactions with parameter $K<1/3$.  The
lowest--order renormalization equations of a weak backscattering potential
are derived explicitly at finite temperature. A perturbative expression for
the conductance in the presence of a potential with arbitrary spatial
extension is given. It depends on the interactions, but is also affected by
the noninteracting leads, especially for very repulsive interactions,
$K<1/3$. This leads to various regimes, depending on temperature and on
$K$. For randomly distributed weak impurities, we compute the conductance
fluctuations, equal to that of ${\mathcal{R}}=g-2e^2 /h$. While the behavior
of $Var({\mathcal{R}})$ depends on the interaction parameters, and is
different for electrons with or without spin, and for $K<1/3$ or $K>1/3$,
the ratio $Var({\mathcal{R}})/{\mathcal{R}}^2$ stays always of the same
order: it is equal to $L_T/L\ll 1$ in the high temperature limit, then
saturates at $1/2$ in the low temperature limit, indicating that the
relative fluctuations of ${\mathcal{R}}$ increase as one lowers the
temperature.  
\end{abstract}

\draft 


\section{Introduction}
One--dimensional quantum wires provide an interesting opportunity to study
mesoscopic physics in an interacting system. On the one hand, it is well
established theoretically that interactions give rise to unique electronic
properties in one dimension, described by the so called Tomonaga--Luttinger
liquid (TLL) model.\cite{bosonisation} A detailed comprehension of the
remarkable interplay between interactions and disorder has been
achieved.\cite{loiT,giamarchi_loc,apel_rice,kane_furusaki} On the other
hand, little attention has been paid to the role of the contacts which are
known to influence strongly the transport properties of mesoscopic
structures and quantum wires. In this respect, two simplified models were
proposed recently. Either one connects a finite wire to the reservoirs by
tunneling barriers,\cite{fil_fini} leading to the suppression of the
ballistic conductance at low temperature due to the dramatic effect of the
interactions. Alternatively, in the opposite situation, the wire is
perfectly connected to one--dimensional leads,\cite{ines,maslov_pon}
yielding a perfect conductance independent on the interactions of any range
less than the wire length.\cite{local} The latter result contradicts the
conductance reduction by the interactions predicted in a wire without
contacts,\cite{apel_rice,kane_furusaki} and is in agreement with recent
experiments on micron--length quantum wires.\cite{tarucha,yacoby_second} The
perfect conductance can be explained through an extension of Landauer's
approach to the interacting wire.\cite{ines_bis} The reservoirs are taken
into account by the flux they inject which acts as an initial condition for
the equation of motion of the density.\cite{ines} The incident flux is
perfectly transmitted for any range of interactions. But reservoirs inject
electrons which are not the proper modes of the wire. This leads us to
introduce intermediate noninteracting leads so that we can properly identify
the injected and transmitted electronic flux. Under theses circumstances an
incident electron undergoes multiple internal reflections at the contacts
due to change in interactions, leading to a perfect transmission into a
series of spatially separated charges.\cite{ines}

While the ballistic conductance of a quantum wire cannot reveal the TLL
character of the wire, the natural question one asks is if the other
spectacular manifestations of the TLL model, that are the signature of its
non--Fermi--liquid behavior, can be still observed. It is the purpose of
this paper to study these features. Some have already been addressed in the
literature,\cite{ines_nato,maslov_disorder,furusaki_fil_fini,these_annales}
but new interesting effects of the leads will be seen to appear. We will
also discuss recent experiments on quantum
wires.\cite{tarucha,yacoby_second}

Beyond these points, in the present paper we develop a formal framework to
deal with the properties of any inhomogeneous Tomonaga--Luttinger liquid
(ITLL). An ITLL can occur in circumstances more general than contact
effects. For example, a spatially varying effective interaction can be due to
a varying width of the wire or to a nearby gate with a peculiar geometry.
One can also think of another ideal system, the edge states in the
fractional quantum Hall effect (FQHE).\cite{wen_prl} The
ITLL model might be relevant to describe transitions between edges at
different filling, or an edge state connected to a Fermi
liquid.\cite{oreg,inhall,chamon_contacts} This motivates us to investigate
not only electrons with spin but also spinless electrons with external
interacting leads. 

Let us now summarize our main results. A typical feature of a TLL is the
separation of the charge and spin dynamics. Imagine that one injects a spin
polarized flux of electrons into one external lead, and detects transmitted
spin and charge on the second external lead. Since the latter is
noninteracting, one might suspect that charge and spin recombine. But we
show that this is not the case: the charge and spin parts are separated even
in the noninteracting leads.

Such process has been first studied in refs.\onlinecite{ines,ines_nato} for
spinless electrons. The propagation is defined in terms of a quasiparticle
current (corresponding to Laughlin quasiparticle in the FQHE), and to a
superposition of electron-hole excitations in a quantum wire. The basic
buildingblock in the description of this effect is the scattering matrix of
one quasiparticle at the contact between a TLL of parameter $K_1$ and a
second one of parameter $K_2$.\cite{ines,ines_nato,these_annales} The
reflection coefficient is $\gamma=(K_1-K_2)/(K_1+K_2)$, thus is negative
when $K_1<K_2$. Such exotic result indicates an analogy with Andreev\cite{andreev,beenakker_bis}
reflection even for repulsive interactions.

In connection with this Andreev reflection, we are here led to show a second
aspect, the manifestation of proximity
effects.\cite{ines,ines_proc_arcs,these_annales} It is known that a typical
feature of the TLL is the nonuniversal algebraic decay of correlations of
different types, reinforcing the tendency towards a CDW or superconducting
order which cannot be of long range due to the importance of quantum
fluctuations in one dimension. It is important to know the effect of the
finite size and of the leads on the fluctuations. We will show here that
the dominant tendency in the wire extends towards the external leads.

A new aspect of the ``mutual proximity effect'' manifests itself when the
wire has a tendency towards a Wigner crystal ($4k_F$ CDW). This happens
usually for Coulomb interactions\cite{schulz_wigner} but also for very
repulsive short range interactions. For electrons with spin with charge
interaction parameter $K<1/3$, the Wigner crystal dominates the $2k_F$ CDW
at any temperature in an infinite wire, while the inverse holds at
$K>1/3$. With external leads connected to a wire with parameter $K<1/3$, we
show the existence of a crossover temperature depending on position $T_c(x)$
below which the $2k_F$ CDW dominates the Wigner crystal. $T_c(x)$ has a
non--trivial powerlaw dependence on both the distance to the contacts and
the wire length.

The behavior of density correlations can be probed through coupling to a
backscattering potential. The effect of weak impurities on the conductance
of the wire with leads was found to be qualitatively similar to that without
leads,\cite{ines_nato,these_annales,maslov_disorder,furusaki_fil_fini} and
this seems in agreement with experiments.\cite{tarucha} But compared to the
usual TLL, the conductance is affected in a nontrivial way by the contacts,
especially when isolated barriers are
considered.\cite{ines_nato,furusaki_fil_fini} We show here that for
electrons with spin this dependence becomes more crucial for very repulsive
interactions because, as mentioned above, the external leads obscure the
tendency towards the formation of a Wigner crystal at low temperature.

Apart from the specific model we consider, we improve the study of weak
impurities. Usually, the renormalization equations in the presence of one
barrier\cite{kane_furusaki} are derived for both the thermal length,
$L_T\simeq v_F/T$, and the wire length, $L$, infinite: their finite values
are accounted for semi--empirically by introducing $\inf({L,L_T})$ as a
cutoff. Here, the renormalization equations are explicitly derived for
finite lengths $L$ and $L_T$. Besides, they are extended to a backscattering
potential whose total spatial extension is less than $L_T$ and $L$.

Another important question concerns the fluctuations of the conductance in
the presence of a random potential distribution. This question can arise in
experiments even if the mean free path is much larger than the wire length
due to the imperfection and roughness on the boundaries. Here we show that
the conductance is self-averaging at high temperature, more precisely
$Var({\mathcal{R}})/{\mathcal{R}}^2$ is of the order $L/L_T\ll
1$.\cite{remark_arovas} But this ratio saturates at $1/2$ at temperatures
low compared to $T_L=u/L$. This has to be taken seriously into account when
one tries to infer precise powerlaws from the experimental value of $g$.

Let us now give the plan of our paper which does not follow the above
sequence of results. Instead, we separate the paper in two parts: the first
deals with spinless electrons which are also relevant for the FQHE, while we
restore spin in the second part. For spinless electrons, we begin by
describing the formalism and the inhomogeneous Luttinger liquid model. We
discuss the discontinuous interaction case which is suitable to understand
qualitatively the results; it might be relevant for edge states in FQHE, but
is has to be taken with some caution for quantum wires.

We then study the density--density correlation functions for two
reasons. First, in order to show how the dominant tendency in the wire
extends towards the external leads, in analogy with proximity effects. Next,
to investigate the role of backscattering. In particular, the nonlocal
correlation functions at finite temperature and at low frequencies are
computed. Any extended non--random weak potential is then considered, for
which the leading renormalization equation is derived explicitly at finite
temperature. The conductance is expressed perturbatively in the
backscattering potential. We discuss the random--impurity case where the
most important new result concerns the conductance fluctuations.

In the second part dealing with electrons with spin, most of the spinless
treatment can be extended except for two important points. First, the
transmission process is studied explicitly. Secondly, we show in which way
leads affect the tendency towards the Wigner crystal that exists for very
repulsive short range interactions. This induces different regimes for the
correction to the conductance in the presence of a backscattering potential.
Finally, we discuss the experimental results obtained in quantum wires.

\section{Spinless electrons}
\subsection{Model}
In a strictly one--dimensional system with short range interactions between
spinless electrons the low--energy properties can be fully parameterized by
two parameters $u$ and $K$. Without interactions $u=v_F$ and $K=1$.  The
Hamiltonian can then be expressed as
\begin{equation}\label{H}
H=\int\frac{dx}{2\pi }\left[ uK\Pi^2+\frac
uK(\partial _x\Phi )^2\right],
\end{equation}
where $\Phi$ is related to the long-wavelength component of the electron
density through $\rho=-\partial_x\Phi/\pi$, and $\Pi$ is the canonical
momentum conjugate to $\Phi$,
$\left[\Phi(x),\Pi(y)\right]=i\pi\delta(x-y)$. Taking into account the
discreteness of the electrons, Haldane derived the representation of the
total electron density in terms of $\Phi$:\cite{haldane_bosons}
\begin{equation}\label{representation}
\rho=-\frac 1\pi\partial_x\widetilde{\Phi}
\sum_{m=-\infty}^{\infty }c_m e^{2im\widetilde{\Phi}}.
\end{equation}
where $\widetilde{\Phi}(x)=\Phi(x)+k_Fx$. The coefficients $c_m$ obey
$c_{-m}= c_m$. For the noninteracting case only $c_0$ and $c_{\pm 1}$ are
nonzero, however in the presence of interactions higher coefficients appear.
The $c_m$ then can be calculated perturbatively, as done implicitly in
ref.\onlinecite{lrk}. 

Consider now an interacting wire delimited by $[-a,a]$ whose length will
be denoted by $$L=2a,$$ connected perfectly to non--interacting leads. The
global system is described by the Hamiltonian (\ref{H}) with spatially
varying parameters $u(x)$, and $K(x)$.\cite{ines,maslov_pon} We require $u$
and $K$ to be uniform on the external leads, { i.e.} for $|x|>L/2$, taking
values $u_L$ and $K_L$. Even though the most relevant situation of
noninteracting leads corresponds to $u=v_F$ and $K_L=1$, it is interesting
to keep $K_L$ for other possible applications, for instance the
FQHE.\cite{inhall}

We note that the absence of translational invariance in the quantum wire
gives rise to an inhomogeneous chemical potential $\mu(x)$ that can be
incorporated by a translation in $\Phi$.\cite{these_annales} Then one has to
replace, in eq.(\ref{representation}),
\begin{equation}\label{phi0}
-k_Fx\rightarrow \phi_0(x)=-k_Fx+\int^x dx' \,\mu \frac K u.
\end{equation} 

Let us discuss the microscopic arguments for the ITLL model. If one assumes
that the screening does not take place in the same way in the
two--dimensional gas into which the wire opens and inside the wire we can
argue that interactions are described by a function $U(x,y)$ which is not
translationally invariant. But then the electronic momentum is not conserved
anymore in the vicinity of the contacts.  Indeed, this appears when one
bosonizes the interaction Hamiltonian $\int\int U(x,y) \rho(x)\rho(y)$,
using $\rho$ from eq.(\ref{representation}): the exponential terms in $\Phi$
cannot be ignored in general, and those terms violate momentum
conservation. But if $U(x,y)=f(x)f(y)h(x-y)$ where $f$ varies slowly on
scales $\lambda_F$, one can reduce the Hamiltonian to a quadratic
form.\cite{these_annales,ines_bis} Strictly speaking, this condition is
required to find a perfect conductance, as discussed in
ref.\onlinecite{ines_bis}. If this condition is not met, and for not too
strong interactions, the perfect conductance is recovered in the high
temperature limit, but is reduced due to the abrupt change in interactions
at low temperature. For instance, when $u(x)=u$ and $K(x)=K$ are constants
for $|x|<L/2$, exponential terms in $\Phi(\pm a)$ cannot be ignored, and
give rise to effective barriers determined by the jump of the interactions.
Nevertheless, due to the symmetry of the structure, and since one needs $k_F
L\gg 1$ to be able to use the TLL model, one can easily achieve resonances
that suppress the role of these terms.

For quantum wires opening adiabatically into a two--dimensional gas, the
situation of abrupt variations is not realistic; it is adopted for
mathematical convenience, and the results can be be trusted only far from
the contacts compared to $\lambda_F$ where the behavior should not depend on
the profile of variations at the contacts. We shall however look at the
behavior at the contact because it is very similar to that in the strong
tunneling limit of two coupled Luttinger liquids:\cite{chamon_contacts} the
effective parameter at the contact, yielding the local conductance
\cite{ines} and controlling the correlation functions\cite{ines_nato} is
exactly the same as the parameter for the equivalent TLL liquid in
ref.\onlinecite{chamon_contacts}.

The situation for edge states in the fractional quantum Hall effect (FQHE)
might be different. Non--quadratic terms arise from tunneling between edge
states, which is more difficult to achieve due to their spatial
separation. Recently, the model with discontinuous parameters was shown to
be suited to describe transitions between edges with different
filling,\cite{inhall} or the connection between an edge state and a Fermi
liquid.\cite{chamon_contacts} In any case, most of the results we derive
here can be extended to any profile of variations of the parameters.

\subsection{Correlation functions}
It is well known that symmetry breaking phase transitions cannot occur in
one dimension. But the interactions in an infinite wire enhance charge
density or superconducting fluctuations depending on whether they are
repulsive or attractive. The natural question is to know whether such
tendencies persist in the presence of leads.

In the present subsection, we would like to compute the correlation
functions not only to answer this question, but also to explain the
proximity effect,\cite{ines,ines_proc_arcs,these_annales} and to give the
necessary tools for studying the role of backscattering in the next section,
closely related to the tendency towards formation of a charge density wave.

\subsubsection{Correlation functions at finite temperature}
\label{properties}
Let us write the density--density correlation function, using
eq.(\ref{representation}):
\begin{eqnarray}\nonumber
\lefteqn{\left\langle T_\tau\rho(x,\tau)\rho(y,0)\right\rangle  =
-\partial_x\partial_y \left[U(x,y,\tau)+
\phantom{\sum_{m\neq 0}}\right.}\\
\label{rhorho}
&&\left. \sum_{m\neq
0}\frac{c_m^2}{2m^2}X_m(x,y,\tau)e^{2im[\phi_0(x)-\phi_0(y)]}\right],
\end{eqnarray}
where $U$ is given in terms of the fundamental bosonic Green
function\cite{gremark}
\begin{equation}\label{Gtau}
G(x,y,\tau )=\left\langle T_\tau\Phi(x,\tau)\Phi (y,0)\right\rangle,
\end{equation}
as 
\begin{equation}\label{Utau}
U(x,y,\tau )=G(x,x,\tau_0)+G(y,y,\tau_0)-2G(x,y,\tau).
\end{equation}
$\tau_0$ is a cutoff different but in general of the order of $1/E_F$. We
impose on the imaginary time $\tau>\tau_0$, so that $G$ can be computed at
finite temperature, without dependence on $\tau_0$.\cite{these_annales}The
consequence of this cutoff procedure on analytic continuation is elucidated
in appendix \ref{correlation}.

In eq.(\ref{rhorho})  $\phi_0$ is given by eq.(\ref{phi0}), and
\begin{eqnarray}\label{Xm}
X_{m}(x,y,\tau)&=&\left\langle T_\tau e^{
im\Phi(x,\tau)}e^{-im\Phi(y,0)}\right\rangle\nonumber\\
&&=e^{-2m^2U(x,y,\tau)}.
\end{eqnarray}
Note that the first term (in $U$) in eq.(\ref{rhorho}) can be obtained
from the limit $m\rightarrow 0$ of $X_m$.

Instead of writing out explicitly $X_m$ the imaginary part of its Fourier
transform in the low frequency limit $\omega\ll T$ will be of more use:
\begin{equation}\label{imaginary}
{\mathcal{I}}mX_m(x,y,\omega\ll T )\sim-\frac{\omega}{E_F^2} \chi_m(x,y),
\end{equation}
where $\chi_m(x,y)$ is a dimensionless function. The function $\chi_m$ can
be obtained from $\chi_1$ by multiplying $K(x)$ by $m^2$, thus we will often
study $\chi_1$, denoted $\chi$ for simplicity. The time dependence of $U$
being due to the nonlocal part $G(x,y,\tau)$ in eq.(\ref{Utau}), we can
write (for details see appendix \ref{correlation})
\begin{equation}
\chi\left( x,y\right)=\frac{1}{\underline{T}^2 }C(x,y)e^{-2G(x,x,\tau_0
)}e^{ -2G(y,y,\tau_0 )}, \label{chic}
\end{equation}
where 
\begin{equation}\label{underlineT}
\underline{T}=T\tau_0
\end{equation}
and $C$ is a dimensionless time integral, eq.(\ref{Gint}). We can find the
properties of $C(x,y)$ for general smooth variations of $K$ and $u$, and we
have computed it explicitly for any $x,y$ in the discontinuous interaction
case and at finite temperature. By deforming the contour of integration in
the complex plane, the computation is made much easier, especially in the
low temperature limit. In the high temperature limit, more steps are
needed. The results are given explicitly in appendix \ref{correlation}.

In an infinite wire with uniform parameter $K$, $\chi(x,y)$ is a function
of $x-y$, and we denote it by $\chi^{(K)}(x-y)$. It decreases exponentially
at separations larger than $L_T=\pi u/T$,
$\chi^{(K)}(r)=(r/L_T)e^{-Kr/L_T}$, while its local value has the typical
powerlaw in temperature $\chi^{(K)}(r=0)\simeq T^{2(K-1)}$ that diverges in
the zero temperature limit when $K<1$, indicating an enhancement of the
density fluctuations compared to the noninteracting case.

For interaction parameters $u,K$ constant in the bulk of the wire and
reaching their asymptotic values $u_L,K_L$ on the external leads, four
general properties of $\chi$ can be shown. The first and second one hold at
high temperature, where comparison can be made with $\chi^{(K)}$.
\begin{enumerate}
\item 
At $T\gg T_L$, we have
\begin{equation}\label{chibulk}
a-\vert x \vert, a-\vert y\vert \gg L_T\Rightarrow
\chi(x,y)\simeq\chi^{(K)}(x-y)
\end{equation}
where $\chi^{(K)}(x-y)$ is the value of $\chi$ in an infinite wire with
parameter $K$. Thus we recover the translationally invariant behavior for
any $x,y$ far from the contacts compared to $L_T$.
\item 
Again, at $T\gg T_L$, but for any location of $x,y$, we have 
\begin{equation}\label{majoration}
\chi(x,y)\leq \chi^{(K)}(x-y)
\end{equation} for $K_L>K$, while the inverse inequality holds for $K_L<K$. 
\item At any temperature, and for small separations compared to $L_T$,
$\chi$ factorizes:
\begin{equation}
\left| x-y\right| \ll L_T\Longrightarrow \chi
(x,y)\simeq\sqrt{\chi(x,x)}\sqrt{\chi (y,y)}. \label{decouple}
\end{equation}
This is due to the fact that $C(x,y)\sim C(x,x)$ [eq.(\ref{chic})], up to
corrections of order $(\vert x-y\vert/L_{T})^2$.
\item
The local value of $\chi$ is almost constant on segments much less than the
wire length and the thermal length. If we denote
\begin{equation}\label{Lmin}
L_{min}=\min(L,L_T),
\end{equation}
then
\begin{equation}\label{chiloc}
\left| x-y\right| \ll L_{min}\Longrightarrow \chi(x,x)\simeq
\chi(y,y)\sim\chi(x,y).
\end{equation}
These equalities hold up to corrections of the order of $(\vert
x-y\vert/L_{min})^2$.
\end{enumerate}
Let us now specialize to the case of discontinuous parameters. First we
introduce the quasiparticle density $j_r=j+ru\rho$ for $r=\pm$ which are the
right-- and left--going proper modes in a uniform TLL. They correspond to
Laughlin quasiparticle currents in edge states of the FQHE. As we said
before, the basic quantity for understanding the physics is the scattering
matrix\cite{ines,ines_nato,these_annales} of quasiparticles at the contact
of two perfectly connected TLL with parameters $K$ and $K_L$. It acts in the
space $(j_+,j_-)$ and is given by
\begin{equation}\label{scattering}
S=\left( 
\begin{array}{cc}
-\gamma & 1+\gamma \\ 
1-\gamma & \gamma
\end{array}
\right) 
\end{equation}
where $\gamma$ is the reflection coefficient
\begin{equation}\label{gamma}
\gamma=\frac{K_L-K}{K_L+K}.
\end{equation}
The matrix $S$ relates the outgoing flux to the injected flux, defined not
in term of wavefunction amplitudes as in usual scattering approaches, but
directly in terms of the quasiparticle current $j_r$. The local effective
parameter (yielding a local Kubo conductance\cite{ines,these_annales}) is
given by $K_1$ multiplied by the transmission
coefficient,\cite{ines,ines_proc_arcs}
\begin{equation}\label{Ka}
K_a=K_L(1-\gamma)=K(1+\gamma).
\end{equation}

Let us first discuss the high temperature limit. It is worth noting that a
lower (upper) bound to $\chi(x,y)$, completing eq.(\ref{majoration}), is
given by $\chi^{(K_a)}(x-y)$ for $K_L\geq K$ ($K_L\leq K$). In
eq.(\ref{chic}), the nonlocal part $C(x,y)$ is given by eq.(\ref{CF1}) in
the limit $T\gg T_L$, while it simplifies to a constant [eq.(\ref{Clow})] in
the limit $T\ll T_L$, in accordance with property (\ref{decouple}). The
Green function $G$ is given by eq.(\ref{apGz}). Instead of writing
explicitly the factor $e^{-2G(x,x,\tau_0)}$ that enters in eq.(\ref{chic}),
we use the local value of $\chi$. This is equivalent as far as the dominant
behavior is concerned because $C(x,x)$ is a slowly varying function of $x$
(cf. appendix \ref{correlation}).  Thus we drop it from
\begin{equation}\label{chiG} \sqrt{\chi(x,x)}\simeq
\frac{1}{\underline{T}}e^{-2G(x,x,\tau_0)}.
\end{equation}
In the high temperature limit $(T>T_L$), the multiple reflections caused by
the change in interactions don't affect the correlation functions due to the
lack of thermal coherence along the wire. Only one reflection is felt within
a thermal length $L_T=u/T$ from the contacts. We now define
\begin{equation}\label{tx}
t_x=\frac{a-|x|}{u\tau_0},
\end{equation}
the time it takes for a quasiparticle emanating at $x$ to reach the closest
contact, measured in units of $\tau_0$.  For $t_x\gg 1$ [eq.(\ref{tx})], we
find
\begin{eqnarray}  \label{Vrenorm}
\sqrt{\chi(x,x)}&\sim& \underline{T} ^{K -1}(\tanh {}
\underline{T}t_{x})^{K_a-K}.
\end{eqnarray}
Again, for points far from the contacts compared to $L_T$, this expression
coincides with that obtained in an infinite TLL with parameter $K$, $
\chi\sim\underline{T} ^{2(K-1)} $. At the contacts, we find
$\chi(a,a)\simeq\sim\underline{T} ^{2(K_a-1)}$ [cf eq.(\ref{chia})] (this
amounts to setting $t_x=1$ in eq.(\ref{Vrenorm}), using $\underline{T}\ll
1$).

In the limit of temperatures low compared to $T_L$, the multiple reflections
affect the correlation functions, thus the external leads with parameter
$K_L$ determine the temperature dependence. But there is a nontrivial
dependence on the wire length as well as on the distance of the barrier from
the contacts:
\begin{eqnarray}  \label{integration}
\sqrt{\chi(x,x)}\simeq& \underline{T} ^{K_L-1}\underline{T}_L^{K_a-K_L}t_{x} ^{K_a-K}.
\end{eqnarray}
This gives the behavior for $t_x\gg 1$. $\chi(a,a)$ can be obtained by
setting $t_x=1$ in this expression.

Remember that one obtains $\chi_m(x,x)$ from $\chi(x,x)$ (in both
temperature limits) by multiplying both $K$ and $K_L$ by $m^2$, which leaves
$\gamma$ [eq.(\ref{gamma})] unchanged. Clearly, the $2mk_F$ CDW is less
important than the $2k_F$ one. Further, if electrons in the external leads
are non interacting, $K_L=1$, the $2mk_F$ CDW are suppressed in the
zero-temperature limit because $\chi_m\sim T^{2(m^2K_L-1)}=T^{2(m^2-1)}$
[eq.(\ref{integration})].

\subsubsection{Proximity effect}
Let us first recall an exotic phenomena emanating from this model, the
analogy with Andreev reflection.\cite{ines,ines_proc_arcs,these_annales}
\paragraph{Andreev reflection} The reflection coefficient of a
quasiparticle from a TLL with parameter $K_1$ incident on another one with
parameter $K_2$ is given by $\gamma$, eq.(\ref{gamma}), and is {\em
negative} for $K_1<K_2$, i.e.  a partial quasi-hole is reflected back. This
is similar to Andreev reflection of an incident electron on a
normal--superconductor interface. The similarity is the closest for $K_1=1$
and $K_2>1$, because the wire has a tendency towards superconductivity, but
it holds more generally even for repulsive interactions, for instance for a
quasiparticle in the wire with $K<1$ incident on the noninteracting lead.

It is worth noting that in the case of two semi--infinite wires with $K_1$
and $K_2$ as parameters, and $K_1<K_2$, the local Kubo conductance at the
interface is given by the transmission coefficient from $1$ to $2$
multiplied by $K_1$, i.e. by $g_a=K_a$. This verifies $K_1\leq g_a\leq
2K_1$.\cite{ines,ines_proc_arcs,these_annales} This is very similar to the
inequalities for an N-S interface, if one interprets $K_1$ as the
conductance (Kubo) of the normal side, $g_N\leq g_{N-S}\leq 2
g_N$.\cite{beenakker}

Finally, let us comment on a recent paper\cite{sandler_andreev} based on the
model in ref.\onlinecite{chamon_contacts} of two TLLs connected by a
tunneling term. The bosonized theory was associated with conformal field
theories to describe quasiparticles in term of soliton
states. Interestingly, in the strong tunneling limit, the scattering matrix
for these states at the interface turns out to be the same as
eq.(\ref{scattering}). This gives a stronger foundation of the quasiparticle
scattering scheme we introduced previously.\cite{ines}
\paragraph{Proximity effect}
The reflection at the contact gives rise to a proximity effect for both
superconducting or CDW
correlations.\cite{ines,ines_proc_arcs,these_annales,maslov_proximity} This
can be seen computing the correlation functions on the external leads. In
particular, when $T\gg T_L$, there is no coherence between the endpoints of
the wire, and it is sufficient to consider half the system. The wire and one
lead have a symmetric role; it is sufficient to permute $K$ and $K_L$ in
eq.(\ref{Vrenorm}) in order to find $\chi$ for $\vert x\vert>a$, and to take
$t_x=(\vert x\vert-a)/u_L$. By the way, this situation is relevant for the
FQHE, if two edge states with different filling are connected.\cite{inhall}

At $T<T_L$,  instead of eq.(\ref{integration}) inside the
wire, outside the wire one has 
$$\sqrt{\chi(x,x)}\simeq\underline{T} ^{K_L-1}(\underline{T}_Lt_{x})
^{K_a-K_L}$$ up to $t_x T_L\sim 1$. Beyond a distance of the order of
$L$ in one external lead, we recover the simple law $T^{K_L-1}$.

For clarity, we discuss the consequence for noninteracting leads,
$K_L=1$. Then if $K<1$, the density--density correlations in the bulk are
similar to those in an infinite wire, but are reduced when one goes to the
contacts because $t_x^{\gamma K}$ decreases for $\gamma >0$. Besides, this
enhancement extends in the external leads up to a distance $L_{min}/K$
[eq.(\ref{Lmin})].\cite{ines_proc_arcs,these_annales} For $K>1$, the pairing
correlation function, which can be obtained from $\chi$ if one replaces $K$
by $1/K$ and $K_L$ by $1/K_L$, is enhanced up to a distance $KL_{min}$. This
increases with $K$, i.e. where interactions are more attractive. This is
reminiscent of the proximity effect.\cite{beenakker}

In general, we can show that the proximity effect extends up to the minimum
length scale at hand. We must note that this holds even for a smooth profile
of $u$ and $K$, as for instance the properties of $\chi$ exposed before are
general.

\subsection{Backscattering by non--random impurities}
We study now the role of impurities in the wire.  The conductance with
impurities was shown to depend on the interactions and is generally affected
by the external
leads.\cite{ines_nato,maslov_disorder,furusaki_fil_fini,these_annales,ines_proc_arcs}
The main goal of this part is to give a general scheme to treat the effect
of a weak backscattering potential $V(x)$ in a non--translational invariant
system and at finite temperature.

\subsubsection{Renormalization Equations}
The coupling of the conduction electrons to impurities with potential $V(x)$
is $\int dx V(x)\rho(x)$, where $\rho$ is given by
eq.(\ref{representation}). It is then convenient to do an integration by
parts giving \cite{ines_resonance}
\begin{equation}\label{nouvelle}
H_{imp}=\int \rho(x)V(x)=\sum_{m=-\infty,m\neq 0 }^{\infty }\int dx \frac
{c_mV'(x) }{2i\pi m}e^{2im\widetilde{\Phi}(x)}.
\end{equation}
The forward scattering term $-\partial_x\Phi V(x)$ is absorbed by a
translation of $\Phi$ by
\begin{equation}\label{translation}
\int^x dx'\frac{K}{u}V(x')
\end{equation}
which has to be included in the scalar function $\phi_0$,
eq.(\ref{phi0}).

 Note that the impurity Hamiltonian (\ref{nouvelle}) is similar to that
considered usually if we replace our $V'(x)$ by $-2imV(x)/(c_mu\tau_0)$. For
instance, if $V$ has the form of a kink, it corresponds to what is commonly
treated as one barrier.

In order to derive the renormalization equations, the exact partition
function $Z$ at finite temperature is expanded in terms of $V'$:
\onecolumn
\begin{eqnarray}
Z&=&\sum_{n} \frac{(-1)^n}{n!}\int \ldots \int
\sum_{\sum_jm_j=0}\prod_{j=1}^nc_{m_j}V'\left( x_j\right) e^{2im_j\phi
_0\left( x_j\right) }\exp \left( \sum_{i\neq j}m_im_jU_{ij}\right),
\label{Zprem}
\end{eqnarray}
where $U_{ij}=U\left( x_i,x_j,\tau _i-\tau _j\right) $ is given by
eqs.(\ref{Gtau},\ref{Utau}), and the integration runs over $x_i$ and
$\tau_i$.  $Z$ describes a neutral gas of integer charges restricted to a
cylinder whose perimeter is $\beta$, and whose height is determined by the
spatial extension of $V'$, reducing to a circle when $V'$ is local. The
charges interact via $U (x,y,\tau )$, eq.(\ref{Utau}). This is not the
Coulomb interaction but an infinite series of logarithmic terms related to
the transmission process discussed earlier (see
eq.(\ref{apGz})).\cite{these_annales} The renormalization procedure is
implemented by increasing the cutoff to $\tau_0(l) =\tau _0e^l$, where
$\tau_0$ is the bare cutoff $1/\Lambda$, and modifying the parameters in
order to keep $Z$ invariant.  It is immediate to derive the
leading--order equation for $V$. According to eq.(\ref{Utau}), the cutoff
appears in the local part of $U$. The change of $U(x,y,\tau)$ due to a
change of the cutoff by $d\tau_0$ is
$$\left[\frac{dG}{d\tau}(x,x,\tau_0(l))+
\frac{dG}{d\tau}(y,y,\tau_0(l))\right]d\tau_0(l),$$ The exponential of such
a term for a pair $x_i,x_j$ factorizes and, once the global neutrality is
used, can be absorbed separately into $V'(x_i)$ and $V'(x_j)$. This gives
the leading-order flow equations for $V(x)$ {\em explicitly at finite
temperature}
\begin{equation}
\frac{dV'_m\left( x,l\right) }{dl}=V'_m\left( x,l\right) \left( 1+2m^2\frac{dG
}{dl}(x,x,\tau _0e^l)\right). \label{dVmsurdlx}
\end{equation}
Note that in the limit of zero temperature, an infinite uniform wire with
parameter $K$, and a purely local $V$, the Fourier transform of this
equation yields the well-known flow equation for $V(2mk_F)=V_m$,
$dV_m/dl=V_m(1-m^2K)$,\cite{kane_furusaki} provided $k_F$ is independent on
the cutoff as we required.

\twocolumn \narrowtext In general, one expects $V(x)$ to renormalize the
interactions. In the extreme case of a local barrier, it was shown by
integrating out degrees of freedom away from the barrier that the
interactions are not renormalized.\cite{kane_furusaki} We can both recover
and generalize this result in a different way, by using the expansion above
at finite temperature and in the finite wire: new interaction terms $W$
between the charges are generated by the renormalization, but they decay
faster than $U $ at long time scales. For instance, in the zero-temperature
limit, and for a barrier in the center of the wire, $U=K\log \tau$ is
corrected by $W(\tau,T=0)=K'(\log\tau)/\tau$, where $K'$ is renormalized by
$V$. At finite temperature, we have $W(\tau,T)\leq W(\tau,0)$, i.e. $W$ is
still decaying faster than $U$.\cite{these_annales}

Let us now discuss a more extended potential than one barrier, but whose
total extension obeys $d\ll L_{min}$ [eq.(\ref{Lmin})]. It turns out that
when one goes to a cutoff such that $u\tau_0(l)>d$, the partition function
(\ref{Zprem}) becomes identical to that of a local
barrier.\cite{these_annales} This is related to the property
(\ref{chiloc}). Then it is tempting to assume that the effect of such a
potential would be similar to that of a local barrier at length scales
larger than the potential extension $d$. But one has to check that the
renormalization from the bare cutoff up to $d$ does not modify the local
interaction parameter, i.e. $K(x)$ at points where $V'(x)\neq 0$. This would
induce non translational invariant effective interactions even if the bare
$K$ is uniform. We can for instance show that $\phi_0$ [eq.(\ref{phi0})] is
renormalized by a complicated complex function, whose equation is rather
lengthy to write. Thus we think this point needs a more thorough study.

Finally, it is worth writing the next--order corrections to
eq.(\ref{dVmsurdlx}) in the case where the partition function becomes
equivalent to that of a barrier and therefore the integrations over $x_i$
can be done explicitly in eq.(\ref{Zprem}). For given $m$, one has to add to
this equation
\begin{equation}
\sum_{m_1+m_2=m} V_{m_1}V_{m_2}
\label{nextleading}
\end{equation}
up to nonuniversal prefactors depending on $m_1,m_2$.\cite{these_annales}

If we ignore renormalizations of the interaction, as is surely correct for a
local barrier or for a weak enough extended potential, the equation
(\ref{renormalisationspin}) can be integrated straightforwardly. In the
absence of an external energy scale, the unique limitation on increasing the
cutoff comes naturally from the fact that one has to put at least two
charges on the same cylinder of radius $\beta $, thus we stop at $\tau
=\beta /2$. Then the renormalized potential is simply
\begin{equation}\label{Vrenormalized}
V'(x,\beta/2)=\sqrt{\chi(x,x)}V'(x)
\end{equation} 
where $\chi(x,x)$, eq.(\ref{chiG}), has been studied in the previous
subsection. 

One can also consider the Fourier transform $V(k)$ of the potential. If the
extension of $V$ is less than $L_{\min}$, we can use eq.(\ref{chiloc}) to
factor out $\chi$ from the integral, thus allowing us to recover the
dominant term of the renormalized Fourier component of $V$:
\begin{equation}\label{Vm}
V_m=\int dx V'(x) e^{2im\phi_0(x)},
\end{equation}
by simple multiplication by $\sqrt{\chi_m(x,x)}$. Note that whenever
$\phi_0$ [eq.(\ref{phi0})] can be replaced by $k_Fx$, $V_m= -2imk_F
V(2mk_F)$. However the nonlocality of $\chi_m$ has to be taken into account
when one looks to the next leading term.\cite{ines_resonance}

The derivation above is valid for any profile of parameters. In case $K$
varies from its external value $K_L$ towards a plateaus at $K$ inside the
wire, we can show that the dependence of $\chi(x,x)$ on $T$ is
monotonic. Its monotony is determined for any $x$ by the sign of $1-K$ at
$T>T_L$, thus is similar to that in an infinite wire. The renormalized $V'$,
Eq.(\ref{Vrenormalized}) obtained at a given temperature goes up (down) when
temperature is lowered for $K<1$ ($K>1$). At $T<T_L$, the $T$ dependence is
controlled by the sign of $1-K_L$, in particular $V'$ saturates for $K_L=1$
at $T<T_L$.

\begin{figure}[htb]
\epsfig{file=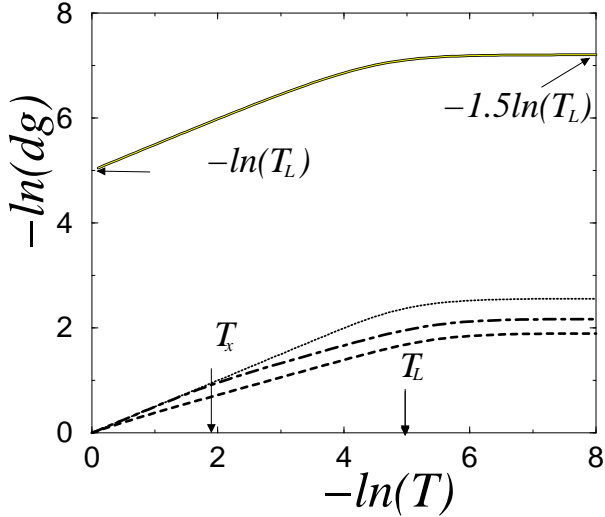,width=9cm}
\caption{The three lower curves show the renormalized barrier strength,
whose square yields the dimensionless reduction to the conductance
$\mathcal{R}$, scaled by its bare value. The dotted, dashed and dot--dashed
curves correspond respectively to a barrier at the center of the wire, at
the contact and at an intermediate point $x$. $T$ is in units of $1/\tau_0$.
We chose $\tau_0= T_L/148$, thus $-\ln (T_L\tau_0)=5$, and $K_\rho=0.5$,
$K_\sigma=1$. The same curves hold for spinless electrons if one replaces
$K_\rho$ by $K$, but one has to multiply the $log\mathcal{R}$ by $2$.  There
is a crossover at $T_L$ from a powerlaw controlled by $K_\rho$
($K_{a\rho}=2K_\rho/(1+K_\rho)$) at the center (at the contact) to a
plateau. For a barrier at $x$, there are two crossovers, one at
$T_x=u_\rho/t_x$ from $T^{K_\rho-1}$ to $T^{K_{a\rho}-1}$, followed by a
saturation occurring at $T_L$. The continuous upper curve corresponds to
$l_e\left\langle \mathcal{R}\right\rangle/(u\tau_0)$ for an extended
Gaussian distribution, where the powerlaw at $T<T_L$ is governed by $K_\rho$
since the interactions are repulsive. The $2k_F$ backscattering dominates
the $4k_F$ one for any $T$ since $K_\rho=0.5>1/3$. }
\label{f:allcourbes}
\end{figure}

In the case of discontinuous parameters, the explicit value of $\chi$ is
given by eqs.(\ref{Vrenorm}) and (\ref{integration}) for the high and low
temperature regime. The renormalized $V$, depending on the location of the
barriers, is shown in fig.\ref{f:allcourbes} for the case $K_L=1$, $K<1$. We
point out that these curves are not inferred from cutting the scaling at $T$
or at $T_L$, but result from the renormalization at finite temperature and
for the finite wire. Note that the saturation at $T<T_L$ occurs only for
$K_L=1$ and for $m=1$.\cite{ines_nato,maslov_disorder}

\subsubsection{Conductance with non--random impurities}
Let us first recall that the conductance of a wire without impurities
$g=e^2/h$ is independent of interactions if the external leads are
noninteracting. The conductance can be related to the transmission, which
turns out to be perfect, thus generalizing Landauer's approach to an
interacting wire.\cite{ines,ines_bis} One can also use the Kubo
formula,\cite{maslov_pon} provided the external leads are noninteracting,
and one is limited to the linear and stationary regime. We have to stress an
important point: the external field has to be used when interactions are
included exactly in the Hamiltonian, and not the internal electric field
modified by the interactions. This is clear for instance through the
equation of motion method we have developed in ref.\onlinecite{ines} to
study transport, for which the {\em external field} forms a source term.

The same conditions to justify the Kubo formula hold with impurities, if one
assumes that the reservoirs impose an external bias $\Delta V$ (that can be
maintained in the noninteracting leads) independently of the current. Then
the stationary current is given by
\begin{eqnarray}
j(x)&=&\lim_{\omega\rightarrow 0}\int \sigma(x,y,\omega)E(y,\omega)\nonumber\\
&&=\sigma(x,y)\Delta V
\end{eqnarray}
where $\sigma(x,y,\omega)$ is the nonlocal conductivity, related to the
Fourier transform of $G$ [eq.(\ref{Gtau})] computed now with the total
Hamiltonian $H+H_{imp}$, and continued to real frequency,
\begin{equation}\label{sigmaG}
\sigma(x,y,\omega)=-\frac{2i\omega}\pi G(x,y,\omega) .\end{equation} The
uniformity of the current as well as time reversal symmetry require the zero
frequency limit of $\sigma(x,y,\omega)$ to be
uniform.\cite{kane_serota,ines,these_annales} Thus
\begin{equation}\label{gsigma}
g=\frac{e^2}h\sigma(x,y).\end{equation}. 

If the external leads are interacting, one cannot impose an external value
of the potential, rather the potential is now renormalized by the
interactions.\cite{local,ines_bis} Nevertheless, the spatial separation of the two
edges in a Hall bar can lead to different couplings with reservoirs as
discussed in \onlinecite{kane_fisher_contacts,chamon_contacts}, and it is
also possible to measure locally the potential on each edge. Thus it is
still of interest to let the external parameter $K_L$ arbitrary.

In appendix \ref{dyson}, we derive a novel exact Dyson equation for the
conductivity in a non--translationally--invariant system. This is very
suitable to write the perturbative expression for the
conductance\cite{these_annales}
\begin{equation}\label{gimp}
g_{imp}=\frac{e^2}h\left(1-\mathcal{R}\right),
\end{equation}
where $\mathcal{R}$ is given by
\begin{equation} \label{RsommeRm}
{\mathcal{R}}=\frac{K_L}{2\pi}\sum_{m=1}^\infty c_m^2 {\mathcal{R}}_{m},
\end{equation}
and ${\mathcal{R}}_{m}$ is the contribution from backscattering of $m$
electrons,
\begin{eqnarray} \label{eqRmnew}
{\mathcal{R}}_m& =&\int\-\-\-\int \chi_m(x,y)\frac{V'(x)V'(y)}{(uk_F)^2}\cos
2m\left[\phi_0(x)-\phi_0(y)\right],
\end{eqnarray}
where $\chi_m$ is defined by
Eqs. (\ref{Utau},\ref{Gtau},\ref{imaginary},\ref{Xm}). Remember that the
forward scattering contribution is included in $\phi_0$
[eq.(\ref{translation})]. We have already obtained general properties of
$\chi(x,y)$. If parameters change abruptly, one can inject the explicit
form (appendix \ref{correlation}) to express the conductance at any
temperature, for sufficient weak potential {\em with any extension}.

If the potential extension is much less than $L_{min}$ [eq.(\ref{Lmin})], we
can use eq.(\ref{chiloc}) to show easily that the dominant term of equation
(\ref{eqRmnew}) reduces to the square of the renormalized potential
\begin{equation}\label{Rlocal}
{\mathcal{R}}_m\simeq\chi_m (x,x)\vert V_m\vert ^2
\end{equation}
where $V_m$ is given by eq.(\ref{Vm}).  As long as $V_1\neq 0$ (i.e. out of
resonance), the dominant term in $\mathcal{R}$ is given by
${\mathcal{R}}_1$. On resonance, the dominant term comes still from $m=1$ or
from $m=2$;\cite{ines_resonance} this depends on the values of $K$ and
$K_L$, as well as the extension and strength of the potential. It appears
that at low temperature, because the $4k_F$ CDW correlation function
acquires a factor $T^{2(4K_L-1)}$ [eq.(\ref{integration}) with $K$ and $K_L$
multiplied by $4$], one must have $K_L>1/4$ for $\mathcal{R}$ to vanish at
zero temperature (for a short enough wire). Remarkably, for $K_L=1$, the low
temperature conductance is still $e^2/h$ on resonance, {\em even for
$K<1/4$}. We will not discuss the resonance in further detail here, even
though one has a variety of behaviors, and we refer to the explicit
expression of $\mathcal{R}$, Eqs.(\ref{RsommeRm}\ref{eqRmnew}), and to the
computation of appendix \ref{correlation}.

Out of resonance, and for discontinuous parameters, $K_L=1$, $\mathcal{R}$
is given by the same curves as those yielding the renormalized $V$ in
fig.\ref{f:allcourbes}. The wire has to be short enough so that $\mathcal{R}$
stays weak enough when reaching the plateaus below $T_L$, otherwise the
perturbative computation breaks down at a temperature above $T_L$. We refer
to ref.\onlinecite{furusaki_fil_fini} for the strong barrier case.

\subsection{Random impurities: conductance fluctuations}
Now we consider random impurities distributed all over the wire. For
simplicity, we limit ourselves to the case of a Gaussian distribution,
$\left\langle V(x)\right\rangle=0$, and
\begin{equation}\label{gauss}\left\langle V(x)V(y)\right\rangle =D\delta
(x-y).
\end{equation} 
By averaging eq.(\ref{eqRmnew}) over disorder, one obtains the leading term
of $\mathcal{R}$ [eq.(\ref{RsommeRm})] coming from the $m=1$ contribution
\begin{equation}\label{Rdes}
\left\langle \mathcal{R}\right\rangle =K_L \int_{-a}^a\frac{dx}{l_e(x)}
\chi(x,x),
\end{equation}
where $1/l_e(x)=4D [\phi_0'(x)]^2/(u k_F)^2$. The forward scattering part in
$\phi_0'$ is $KV(x)/u$ [eq.(\ref{translation})], and thus can be dropped
when averaging. Thus the dependence of $l_e$ on $x$ is due to the
inhomogeneity of the interactions. It could also be induced by a
space--dependent disorder strength $D(x)$. Nevertheless, we will assume
$\phi_0'(x)\simeq k_F$ for simplicity, in which case $l_e$ is uniform and is
equal to $D/u^2$. This holds for instance in the bulk of the wire, and thus
this is a plausible approximation whenever the impurities in the bulk
dominate.

Note that we have obtained eq.(\ref{Rdes}) by performing first a double
integration by parts in eq.(\ref{eqRmnew}) and then retaining only the term
with no derivative of $\chi$. The explicit integration of the function
$\chi$ can be done but is tedious. However, the results can be understood
easily.\cite{ines_nato,these_annales} There are two main contributions: one
from the bulk that behaves, for $T>T_L$, as $LT^{2(K-1)}$, the other from
the contacts behaving as $T^{2(K_a-1)}$. For $T<T_L$, up to nonuniversal
constants, these contributions become\cite{ines_nato,these_annales}
\begin{equation}\label{lowmathR}
\left\langle{\mathcal{R}}\right\rangle\simeq
\frac{u\tau_0}{l_e}\underline{T}^{2(K_L-1)}\left[
T_L^{2(K-K_L)-1}+T_L^{2(K_a-K_L)}\right]
\end{equation} 
where $K_a$ is given by eq.(\ref{Ka}).  

Let us discuss the case of noninteracting leads, $K_L=1$. In this case, the
contact contribution dominates for very attractive interactions, $K>(3+\sqrt
17)/2$, and for not too high temperatures (or not too long wires).

Inspecting eq.(\ref{lowmathR}), we see that
$\left\langle{\mathcal{R}}\right\rangle$ saturates at $T<T_L$ at a value
$\ll 1$ for $K>3/2$, but that it increases with wire length for $K<3/2$. In
the latter case, the perturbative computation is valid only for $L<L_{loc}$
where
\begin{equation} \label{eq:Lloc}
L_{loc}\sim u\tau_0 \left(\frac{l_e}{u\tau_0}\right)^{\frac{1}{3-2K}}.
\end{equation} 
This coincides with the localization length inferred by scaling arguments in
\onlinecite{apel_rice,giamarchi_loc}.

 Let us now compute the variance of the
conductance $Var (g)=\left\langle g^2\right\rangle-\left\langle
g\right\rangle^2$. Clearly, it is also equal to the variance of
$\mathcal{R}$ which can be expressed perturbatively, using
eq.(\ref{eqRmnew}), as
\begin{equation}\label{varianceR}
Var({\mathcal{R}})=K_L^2\int_{-a}^a\int_{-a}^a\frac{dxdy}{l_e^2}
\cos^22\left[\phi_0(x)-\phi_0(y)\right] \chi^2(x,y).
\end{equation} 
Here we have used eq.(\ref{gauss}) to perform the average of
${\mathcal{R}}^2$, have dropped the product of two correlation functions for
different pairs of integers, and neglected the role of the derivative in
eq.(\ref{nouvelle}).

 It is interesting to consider first the low temperature limit $T<T_L$,
where a general inequality, valid for arbitrary profile $K(x)$, can be
shown. Using the factorization property (\ref{decouple}) which holds now for
any $x,y$ in the wire because $L\ll L_T$, we can write eq.(\ref{varianceR})
as
\begin{equation}\label{variance}
Var({\mathcal{R}})=\sum_{r=\pm 1}\frac{K_L^2}2 \left|
\int_{-a}^a\frac{dx}{l_e} e^{2i(1+r)\phi _0(x)}\chi (x,x)\right| ^2.
\end{equation}
The term for $r=1$ on the right hand side is less than that for $r=-1$, this
latter is nothing but $\left\langle{\mathcal{R}}\right\rangle^2$
[eq.(\ref{Rdes})]. Thus we have the interesting inequality
\begin{equation}\label{encadrereflection}
\frac12\left\langle \mathcal{R}\right\rangle ^2\leq Var({\mathcal{R}})\leq
\left\langle \mathcal{R}\right\rangle ^2.
\end{equation}
In particular, this shows that
\begin{equation}\label{ratiolow}
\frac{Var({\mathcal{R}})}{\left\langle \mathcal{R}\right\rangle
^2}\simeq\frac{1}2.
\end{equation}
If $\phi_0(x)$ is simply $k_Fx$, this is a very good estimate of
$Var({\mathcal{R}})$ because the integral for $r=1$ in eq.(\ref{variance})
contains a rapidly oscillating function, $e^{2ik_Fx}$, while $\chi$ varies
slowly compared to $\pi/k_F$. Besides, $Lk_F\gg 1$ in order to
validate the bosonization, so that the integral corresponding to $r=1$ in
eq.(\ref{variance}) can be neglected. This justifies eq.(\ref{ratiolow}),
showing that the fluctuations of $\mathcal{R}$ are of the same order as its
average value. In the special case of short range interactions with
discontinuous parameters, $Var({\mathcal{R}})$ is obtained as the square of
eq.(\ref{lowmathR}). 

We now consider the high temperature limit, $T>T_L$. To simplify the
discussion, we restrict ourselves to the case $K<1$, which is probably the
case in a quantum wire. Then the bulk contribution dominates over the
contact effects, so that we can restrict the integral in
eq.(\ref{varianceR}) to $-a/2,a/2$, and therefore $ t_x T\gg 1$. Over this
segment, $\chi$ depends only on $x-y$ ( see eq.(\ref{chibulk})), and we can
write, after a change of variables $x-y=vL_T$,
\begin{equation}
Var({\mathcal{R}})\simeq\frac{L L_T}{l_e^2}\underline{T}^{4(K-1)}I(2k_FL_T)
\end{equation}
where 
\begin{equation}I(\xi)=\int_0^\infty (1+\cos{2\xi
v})\left[C_{K}(v)\right]^2 dv,
\end{equation} 
and $C_K$ is given by eq.(\ref{CKv}).  Since $L\gg L_T$, we have extended
the integral approximatively to infinity. But since we are restricted to temperatures much less than the Fermi energy, we have $\xi=k_FL_T\gg 1$, thus the oscillatory part of $I$ can be
neglected and $I$ tends to a constant.  The ratio
\begin{equation}\label{ratiohighT}
\frac{Var({\mathcal{R}})}{\left\langle{\mathcal{R}}\right\rangle^2}\sim\frac{L_T}{L} 
\end{equation}
is now much less than unity, contrary to the low temperature limit, thus the
relative fluctuations of ${\mathcal{R}}$ are small.  It is worth noting that
one can extrapolate the ratio (\ref{ratiohighT}) obtained for $L\gg L_T$ to
$L\simeq L_T$, to recover the same order of magnitude as
eq.(\ref{ratiolow}) at which $Var({\mathcal{R}})/\left\langle{\mathcal{R}}\right\rangle^2$ saturates
at low temperature, i.e. for $L_T>L$.

\section{Electrons with spin}
Let us take into account the spin of the electrons, which is necessary when
dealing with quantum wires, and to confront the theory with experiments.
This case might eventually also be relevant for two-channel edge states in
FQHE.

\subsection{Bosonization}
The typical feature for interacting electrons with spin is the separation
of the charge and spin degrees of freedom at low energy. The density for
each spin component $\rho_s$ ($s=\pm$ denoting the spin up or down) has a
long-wavelength part that is related to a boson field $\Phi_s$ by
$\rho_s=-\partial_x\Phi_s/\pi$.  The total density of electrons with spin $s$
can be written as eq.(\ref{representation}) where $\Phi$ has now the
subscript $s$.  The charge and spin fields are defined by
$$\Phi_{\rho,\sigma} =(\Phi_{\uparrow}\pm\Phi_{\downarrow})/\sqrt{2}.$$ The
Hamiltonian describing the low-energy properties can be decoupled in a
charge and spin parts, $H=H_\rho+H_\sigma$, where
\begin{equation}\label{Hnuspin}
H_\nu =\int \frac{dx}{2\pi }\left[ u_\nu K_\nu  \Pi_\nu
 ^2+\frac{u_\nu }{K_\nu }\left( \partial _x\Phi _\nu \right) ^2\right]
\end{equation}
for $\nu =\rho,\sigma$. The boson fields are related to the charge and spin
density ($\rho$ and $\sigma$) through ${\sqrt{2}}\partial _x\Phi _\rho
(x)/\pi =\rho$ and ${\sqrt{2}} \partial _x\Phi _\sigma /\pi=
\sigma$. $\Pi_\nu$ is the momentum density conjugate to $\Phi_\nu$. In the
absence of interactions, $K_\nu=1,u_\nu=v_F$.

Recall that an additional term has normally to be added to $H_\sigma$
[eq.({\ref{Hnuspin}})] corresponding to the backscattering of electrons of
opposite spin, $g_{\perp}\int dx \cos{\sqrt8\Phi_\sigma}$. For
spin-invariant interactions, $g_{\perp}$ renormalizes to zero if it is
initially positive, while a spin gap is opened if it is negative. Then
$K_\sigma$ scales respectively to $1$ or zero which are the only values
consistent with SU$(2)$ symmetry.

 The inhomogeneous TLL model is now characterized by $x$ dependent functions
$u_\nu(x)$ and $K_\nu(x)$.  For simplicity, we take noninteracting leads,
otherwise the parameters would be too numerous; indeed this is the most
relevant situation for quantum wires.  Nevertheless, the treatment of
$g_{\perp}$ is not trivial for non--translationally invariant interactions
because $g_\perp$ now is space--dependent.  We have derived the
renormalization equations for this case ,\cite{these_annales} nevertheless
their integration is quite involved.  We can however draw qualitative but
not firm conclusions, restricting ourselves to spin isotropic interactions.

The most difficult case corresponds to attractive interactions. As in the
uniform case, we expect a spin gap to develop, but it is not clear how it
extends spatially and how is it affected by the leads in the low temperature
regime. The way $K_\sigma$ would vary from $1$ on the external leads to
$K_\sigma^*=0$ inside the wire is not clear neither; this is an interesting
problem to solve.

For repulsive interactions, the inhomogeneous $K_\sigma$ is expected to
 renormalize towards a value respecting the SU(2) symmetry
 $K_\sigma^*=1$. Loosely speaking, this is easier to study because the
 leads have $K_\sigma=1$, so that they don't prevent this scaling and rather
 favor it. This seems the most relevant situation for quantum wires, even
 though we can sometimes allow for $K_\sigma$ to take any value.
 
\subsection{The transmission process}
The first issue we consider is the transmission process of an incident
electron. The case of electrons without spin was discussed in
refs.\onlinecite{ines,ines_nato}, and this can be extended easily to
electrons injected from reservoirs.

 Let us imagine that we inject an electron of definite spin in the left
lead. This corresponds to create a kink in both $\Phi _\rho $ and $\Phi
_\sigma $ whose time evolution has to be solved given the initial
conditions. The equations of motion for these fields are decoupled, and
require their continuity at the contacts as well as that of $u_\nu \partial
_x\Phi _\nu /K_\nu $. In particular, both the charge and spin current, $
j_\nu =\sqrt{2} \partial _t\Phi _\nu/\pi $ are conserved. This is because we
neglect interactions that violate the conservation of $j_\rho$ and
$j_\sigma$, corresponding respectively to the umklapp process and the
backscattering of electrons with opposite spin ($g_{\perp}$ is irrelevant
for repulsive interactions, see ref.\onlinecite{these_annales} for the
ITLL).

\begin{figure}[htb]
\epsfig{file=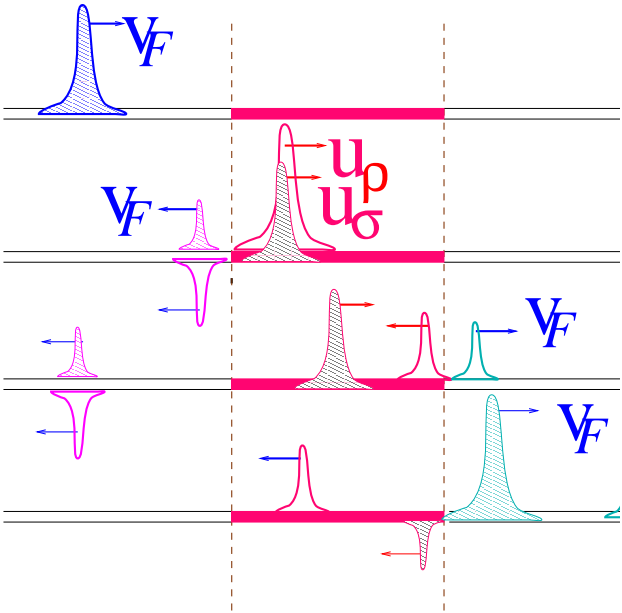,width=9cm}
\caption{Dynamic transmission of an incident electron with spin up. The
charge and spin (hatched) are separated even in the noninteracting leads. As an
example, we consider here $u_\rho>u_\sigma$ , $K_\sigma<1$, $K_\rho>1$.}
\label{f:refspin}
\end{figure}

In view of the above continuity requirements, the charge and spin of the
incident electron are reflected at the first contact with two different
coefficients similar to eq.(\ref{gamma}) with a subscript $\nu=\rho,\sigma$,
\begin{equation}\label{gammanu}
\gamma _\nu =\frac{1-K_\nu}{1+K_\nu}.
\end{equation} 
The transmitted charge and spin propagate at different velocities $u_\rho
\neq u_\sigma $ and thus reach the second contact where they get partially
transmitted at different times $t_\nu =L/u_\nu $. Since the transmitted
charge and spin propagate at the same Fermi velocity in the right
non--interacting lead, they will stay {\em spatially separated}, at a
distance $v_F\left| t_\rho -t_\sigma \right| $ (fig.\ref{f:refspin}). Due to
subsequent internal reflections, we end up with a series of spatially
separated partial charge and spin spikes on the two non interacting
leads. The transmitted spikes correspond to a complicated superposition of
electron-hole excitations. Note that in the relevant case of spin-invariant
repulsive interactions, $K_\sigma=1$, the spin part is not reflected but
gets directly transmitted to the right lead, while the charge undergoes the
multiple reflection process. At times very long compared to $t_\rho $ ($
t_\sigma $), the series of transmitted charge (spin) spikes sum up to
unity. Thus the transmission of an incident spin polarized flux is perfect.

\subsection{"Phase diagram"}
In an infinite wire, interacting electrons with spin have a quite involved
``phase diagram''; here we do not intend to study it entirely when the wire
is connected to leads. We give some qualitative results and then discuss
thoroughly the tendency towards the formation of a Wigner crystal.  As for
spinless electrons, we expect the dominant tendency to extend towards the
leads in analogy with the proximity effect, but the leads can also intervene
in the competition as will be shown later, especially at low enough
temperature.

Let us first discuss the case where superconducting order can take place, as
is the case in an infinite wire with $K_\rho>1$. One has to distinguish
singlet SS, corresponding to the spin gap case, $K_\sigma^*=0$ and triplet
SS, $K_\sigma^*=1$. If we naively take these values for $K_\sigma$ inside
the wire, then we can show that:

For $K_\sigma^*=0$ inside the wire in order to take into account the spin
gap, we can show that the tendency towards singlet super-conductivity holds
for $K_\rho>1$, but the proximity effect towards the external leads shows up
only for $K_\rho>3$. Remarkably, for $K_\rho\gg 1$, an incident electron
with spin up is reflected back into one hole with spin down, and two
quasi-particles of opposite spin, of charge unity and moving at velocity
$u_\rho$ are transmitted, recalling Andreev
reflection. \cite{ines,these_annales}

For $K_\rho>1$, and $K_{\sigma}^*=1$, we can show that triplet
superconductivity develops and extends towards the external leads, but the
Andreev reflection is more subtle to interpret.\cite{these_annales}

In the following, we focus on the situation $K_\rho<1$, $K_\sigma^*=1$, as
it should be for quantum wires, and determine the dominant CDW in the
density--density correlation function.

\subsubsection{Density--density correlations}
Let us compute the density-density correlation function. In order to express
the density, we can superpose eq. (\ref{representation}) for the spin up or
down, thus with the additional index $\nu=\rho$ or $\sigma$, then express
$\Phi_\nu$ as function of $\Phi_\rho,\Phi_\sigma$. This yields
\begin{equation}\label{rhospin}
\rho(x)=-\partial_x\left[\widetilde{\Phi}_\rho+\sum_{m_\rho\neq
0,m_\sigma}\frac{1}{2im_\rho} e^{i\sqrt{2}\left( m_\rho \widetilde{\Phi}
_\rho +m_\sigma\Phi _\sigma \right) }\right],
\end{equation}
where the sum runs over integers $m_{\rho}$ and $m_{\sigma}$ of the same
parity,\cite{kane_furusaki} and
$\widetilde{\Phi}_\rho=\Phi_\rho-k_Fx$. Normally only $m_\rho=\pm m_\sigma$
are allowed. But the other harmonics with $m_\rho\neq m_\sigma$ are
generated by renormalization in the induced density that develops as a
response to an external perturbation.\cite{these_annales}

The fermionic correlation functions can be expressed through the bosonic
correlation function $U_\nu$ (and its dual) for $\nu=\rho,\sigma$ 
as in eq.(\ref{Utau}) with $u(x)$ and $K(x)$ indexed by $\nu$. Then the
density--density correlation function is similar to eq.(\ref{rhorho}) if one
adds the subscript $\rho$ to $U$ in the first line, and $X_m$ is replaced by
$X_{m_\rho,m_\sigma}$ obtained from eq.(\ref{Xm}) by the following
substitution
\begin{equation}2m^2U\rightarrow 
m_\rho^2U_\rho+m_\sigma^2U_\sigma\label{receipe}. \end{equation} 
Thus $$X_{m_\rho,m_\sigma}=\sqrt{X_{m_\rho}X_{m_\sigma}},$$ where $X_{m_\nu}$
is the analogous of $X_m$ without spin with $u$ and $K$ indexed by $\nu$.
The imaginary part of the Fourier transform of $X_{m_\rho,m_\sigma}$ at low
frequencies, $\chi_{m_\rho,m_\sigma}$ is defined as in
eq.(\ref{imaginary}). It cannot be factored unless the spin and charge
velocities are equal. For studying the dominant tendency,
it is sufficient to consider its local values. Up to a slowly varying
function of $x$, we have
\begin{equation}\label{factorspin}
\chi_{m_\rho,m_\sigma}(x,x)=
\sqrt{\chi_{m_\rho}(x,x)}\sqrt{\chi_{m_\sigma}(x,x)},
\end{equation} 
where $\chi_{m_\nu}(x,x)$ is exactly that computed without spin, with the
additional index $\nu$, in particular there are two thermal lengths
associated with the different charge and spin velocity $L_{T,\nu}=u_\nu/T$,
two temperatures $T_{L,\nu}=u_\nu/L$, and two times (in units of $\tau_0$)
$$t_{x,\nu}=\frac{a-\vert x\vert}{u_\nu \tau_0}$$ for a spin ($\nu=\sigma$)
or charge ($\nu=\rho$) excitation to go from $x$ to the closest contact.

The calculation can be carried out for any parameters for charge and spin,
on the leads or inside the wire, but let us specify eq.(\ref{factorspin})
for non interacting leads, and for $K_\sigma=1$. In the high temperature
limit, using eq.(\ref{Vrenorm}),
\begin{eqnarray}\label{chihighspin}
\chi_{m_\rho,m_\sigma}(x,x)&=&
\underline{T}^{m_\rho^2K_\rho+m_\sigma^2-2}(\tanh {}
\underline{T}t_{x})^{m_\rho^2\gamma_\rho K_\rho}.
\end{eqnarray}
In particular, at a distance much greater than $L_{T,\rho}$ from the
contacts, $\chi_{m_\rho,m_\sigma}(x,x)$ is the same as that in an infinite
wire
\begin{equation}\label{chispinbulk}
\chi_{m_\rho,m_\sigma}(x,x)\sim\underline{T}^{m_\rho^2K_\rho+m_\sigma^2-2}.
\end{equation}
 For $m_\rho^2K_\rho+m_\sigma^2<2$, the $(m_\rho,m_\sigma)$ component of the
density--density correlation function is enhanced compared to the
noninteracting case. 

In the low temperature limit, using eq.(\ref{integration})
\begin{equation}\label{chilowspin}
 \chi_{m_\rho,m_\sigma}(x,x)=\underline{T}
 ^{m_\rho^2+m_\sigma^2-2}\underline{T}_{L\rho}^{-m_\rho^2\gamma_\rho}t_{x,\rho}
 ^{m_\rho^2\gamma_\rho K_\rho}.
\end{equation} 

Note that the superconducting correlation functions can be obtained in a
similar fashion, one has to distinguish the triplet from the singlet
superconducting tendency, and we refer to \onlinecite{these_annales} for more
details.
 
\subsubsection{The suppression of the Wigner crystal}
 In an infinite wire, the $4k_F$ CDW, corresponding to Wigner crystal,
dominates the $2k_F$ CDW for very repulsive interactions, more precisely for
$K_\rho<1/3$ and $K_\sigma=1$ as one can inspect using
eq.(\ref{chispinbulk}).  We show here that the leads can induce also a
proximity effect in the opposite sense, by suppressing the Wigner Crystal at
low temperature.  This tendency is even more important for long--range
Coulomb interactions to which case one could extend qualitatively the
results obtained here.

Our study without spin has shown that the enhancement of a $2k_F$ CDW
persists with leads. We have also observed that the $2mk_F$ CDW are
suppressed at low temperature due to the $T$ dependent term $T^{2(m^2-1)}$
in $\chi_m$; but this effect intervenes only when the backscattering
potential is studied on resonance, because the $2k_F$ CDW dominates $2mk_F$
for $m\geq2$.

\begin{figure}[htb]
\epsfig{file=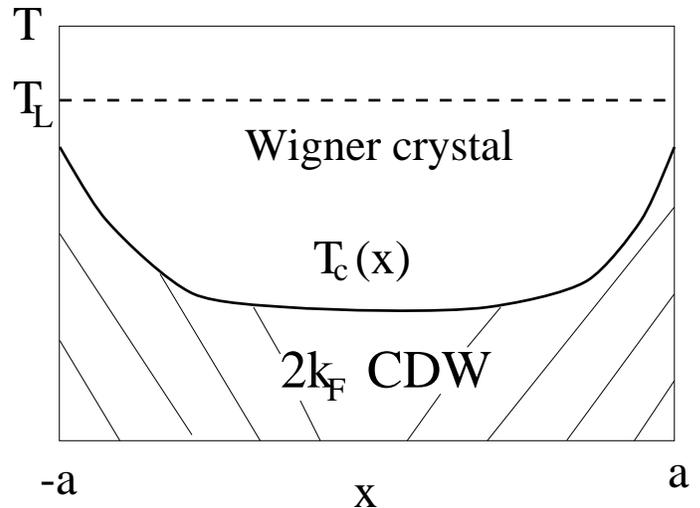,width=9cm}
\caption{The case $K_\rho<1/5$: the Wigner crystal is suppressed below the
crossover temperature $T_c(x)$, eq.(\ref{Tcl}). }
\label{fg:KlessT}
\end{figure}

\begin{figure}[htb]
\epsfig{file=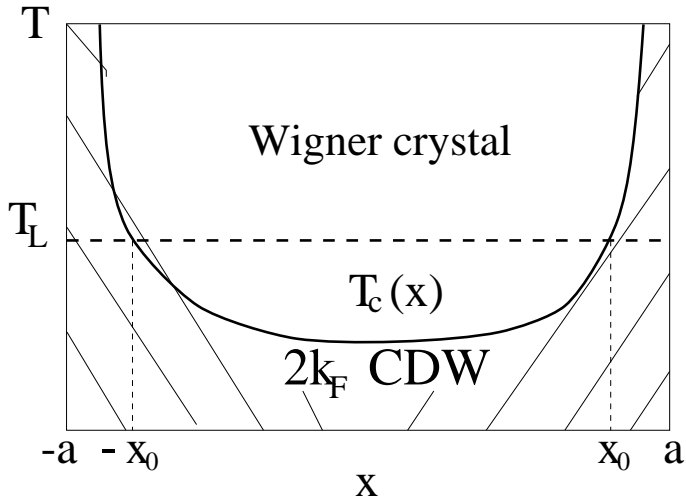,width=9cm}
\caption{The same as fig.\ref{fg:KlessT}, but with $1/5<K_\rho<1/3$.}
\label{fg:Kinterm}
\end{figure}

With spin, one has to compare the $2k_F$ to the $4k_F$ CDW, thus the
correlation functions $\chi_{m_\rho,m_\sigma}$ corresponding respectively to
$(m_\rho,m_\sigma)=(1,1)$ and $(2,0)$.  The competition is much less trivial
than without leads and is \cite{remarkkf} discussed in detail in appendix
\ref{appwigner}. Here we summarize the results, illustrated by
figs.\ref{fg:KlessT} and \ref{fg:Kinterm}. 

We will focus on the $K_\rho<1/3$ case, since for $K_\rho>1/3$ the $2k_F$
CDW is dominant at any temperature. Then contrary to the infinite wire where
the $4k_F$ dominates, we show here that there is a crossover temperature
$T_c(x)$ from the $4k_F $ CDW to the $2k_F$ CDW as $T$ is lowered below
$T_c(x)$. $T_c(x)$ decreases monotonically when $x$ varies from one contact
to the center of the wire where it reaches its minimum value, interestingly
given by the following power of $T_L$:
\begin{equation}\label{Tc0}
\underline{T}_c(0)=\underline{T}_L^{\frac 3 2(1-K_\rho)}.
\end{equation}
One can check that $T_c(0)<T_L$ because $K_\rho <1/3$. Indeed $T_c(x)$
saturates at $T_c(0)$ for all $x$ such that $a-\vert x\vert \sim a$,
i.e. $t_x T_L\sim 1$.
Thus for any $T<T_c(0)$, the Wigner crystal is suppressed all over the
wire and the dominant tendency is towards the $2k_F$ CDW.

In order to discuss the case $T>T_c(0)$ we shall investigate in detail the
behavior close to the contacts. There is a second particular value other
than $1/3$ that arises: all the correlation functions at $\pm a$ are
analogous to those in the center of the wire with $K_\nu$ replaced by
$K_{\nu a}=2K_\nu/(1+K_\nu)$ (the analogue of eq.(\ref{Ka}) with the
supplementary index $\nu =\rho,\sigma$). When $K_\sigma=1$, $K_{a\sigma}=1$,
the limiting value $K_{a,\rho}=1/3$ corresponds to $K_\rho=1/5$. In
particular, as long as $K_\rho >1/5$ one has $K_{a,\rho}>1/3$, thus the
$2k_F$ CDW dominates the $4k_F$ CDW at the contacts at any temperature.

This yields different expressions for $T_c(x)$ depending on $K_\rho$:
\begin{itemize}
\item For $1/5<K_\rho<1/3$, it is given by eq.(\ref{TcKmore}), and
diverges at $\pm a$; reassuringly, we find that $T_c(x)$ is continuous at
the point $x_0$ [eq.(\ref{tx0})] where it reaches $T_L$ from below or above
(fig.\ref{fg:Kinterm}).
\item 
 For $K_\rho<1/5$, $T_c(x)$ is given by Eq.(\ref{Tcl}).  $T_c(x)$ is below
$T_L$ for any $x$, and reaches its maximum value at $a$; $T_c(a)$ is similar
to eq.(\ref{Tc0}) where $K_\rho$ is replaced by
$K_{a\rho}$. (fig.\ref{fg:KlessT})
\end{itemize}

If we consider now any $K_\rho<1/3$ and a given temperature in the range
$[T_c(0),T_c(a)]$ ($T_c(a)$ is infinite for $K_\rho>1/5$), we see that the
dominant tendency switches from the $4k_F$ to the $2k_F $ CDW when one goes
from the center to the contacts.  The point $x$ at which this transition
occurs is a temperature dependent function $x_c(T)$ that is the inverse of
$T_c(x)$. For $T$ higher than $T_c(a)$ (lower than $T_c(0)$), the $4k_F$
($2k_F$) CDW dominates all over the wire.

To conclude, the external leads {\em suppress the importance of the $4k_F$
CDW} at low enough temperature or close to the contacts, thus preventing the
tendency towards the formation of a Wigner crystal.

\subsection{Backscattering by non--random impurities} 
The role of impurities when spin is taken into account can be treated in a
similar fashion as without spin, by doubling the indices as above,
leading to different powerlaws on temperature. Nevertheless, in view of our
previous discussion, the main additional complication arises from the
competition between the $2k_F$ and $4k_F$ backscattering: the leads
intervene more strongly if $K_\rho<1/3$.

We will first derive briefly the renormalization equations at finite
temperature, then give results for the conductance.

The coupling to impurities is described in terms of $\int \rho(x)V(x)$,
where $\rho$ is given by eq.(\ref{rhospin}). For any potential, the
partition function can be expanded in terms of $V'$ in a similar way as for
the spinless case [eq.(\ref{Zprem})], but with $m_i$ replaced by a couple of
integers $(m_{\rho,i},m_{\sigma,i})$. Then $Z$ describes integer charges
restricted to two neutral cylinders, corresponding to the charge ($\nu =\rho
$) and the spin ($\nu =\sigma $) degree of freedom. The radius of each
cylinder is given by $\beta/\pi$, and its height determined by the spatial
extension of the potential. Only charges on the same cylinder $\nu$ interact
via $U_\nu (x,y,\tau )$, eq.(\ref{Utau}) with the index $\nu$ on any
function or parameter.

Following similar steps as in the spinless case, we can infer the leading
renormalization equation analogous to (\ref{dVmsurdlx}), where $U$ is
replaced according to the recipe (\ref{receipe}),
\begin{eqnarray}  \label{renormalisationspin}
\frac{dV'(x.m_\rho,m_\sigma)}{dl}&=& \left[1-\frac{1}{2}\left(m_\rho^2\frac{
dU_\rho}{dl}+m_\sigma^2\frac{dU_\sigma}{dl} \right)\right]V' .
\end{eqnarray}
For simplicity we have omitted the arguments on the right hand side. In
particular, through $U_\nu$, $V$ now acquires an additional spatial
dependence.

The bare value of $V(x,m_\rho,m_\sigma)$ is zero if these terms are not
present in the initial density representation, eq.(\ref{rhospin}). In order
to illustrate how they are generated, we specialize to the case of one
barrier. Then the cylinders shrink to circles, and the partition function
reads
\begin{eqnarray}\label{Zonebarrier}
Z&=&\sum_n(-1)^n\int_{(n-1)\tau_0 }^{\beta -\tau_0 }{d\tau _n} \ldots
\int_0^{\tau _{2-}\tau_0 }{d\tau _1}\nonumber\\ &&\sum_{\sum_{i}m_{\nu
i}=0}\prod_{i=1}^n V_{m_{\rho i},m_{\sigma i}}\exp\left[{ \sum_{i\neq
j,\nu}m_{\nu i}m_{\nu j}U_\nu\left(\tau_i-\tau_j\right)}\right].
\end{eqnarray}
The sum runs over all $n$-tuples of integers $m_{\rho i}$ and $m_{\sigma_j}$
with total vanishing sum separately, and the bare transform of $V$
\begin{equation}\label{Vmspin}
V_{m_{\rho },m_{\sigma }}=\int V'(x)e^{2im_\rho\phi_0(x)}
\end{equation}
acquires a dependence on $m_\sigma$ during renormalization.  We can show
that the interactions are not renormalized as for the spinless case. The
next--leading term to eq.(\ref{renormalisationspin}) is obtained by the
contraction of one pair of charges on each circle, and is equal
to\cite{these_annales}
\begin{equation}\label{higher_m_spin}
\sum V(n_\rho,n_\sigma)V(n'_\rho,n'_\sigma),\end{equation} up to non universal
prefactors, the sum running over
$n_\rho+n'_\rho=m_\rho$ and $n_\sigma+n'_\sigma=m_\sigma$.
 
For a potential with extension $d$ much less than $L$ and $L_T$, we can show
that the partition function is analogous to eq.(\ref{Zonebarrier}) when a
cutoff larger than $d$ is reached. Again, at the location of $V$ the
interaction parameters might be renormalized before reaching this stage.

If the renormalization of the local effective interactions, as well as next
leading terms can be ignored, eq.(\ref{renormalisationspin}) can be
integrated easily. The maximum cutoff is half the radius of one cylinder,
i.e. $\beta/2$, where we get the renormalized parameters\cite{remark_m}
 $$V'(x;m_\rho,m_\sigma)=V'(x)\chi_{m_\rho,m_\sigma}(x,x).$$
$\chi_{m_\rho,m_\sigma}(x,x)$ has been studied previously, and is given by
eq.(\ref{chihighspin}) (respectively (\ref{chilowspin})) in the high (low)
temperature limit.

 For $K_\sigma=1$, $K_\rho<1$, and for any $x$, $V'$ is enhanced when the
 temperature is lowered, and is less enhanced when one gets closer to the
 contacts because the leads moderate the repulsive interactions. If the
 potential is weak enough or the wire is short enough, we can perform the
 renormalization at temperatures below $T_L$. The renormalized $V'(x,1,1)$
 obtained at a given $T<T_L$ is the same for all $T<T_L$, but this is not
 the case for all the other couples $(m_\rho,m_\sigma)\neq (1,1)$, for which
 $V'(x,m_\rho,m_\sigma)$ goes down to zero at zero temperature. For
 instance, we refer to our previous discussion of the competition between
 the $2k_F$ and $4k_F$ backscattering contribution.

Now we give the perturbative expression for the conductance.  For a general
weak potential $V(x)$, the conductance can be expressed as in
eq.(\ref{gimp}). $\mathcal{R}$ is similar to eq.(\ref{RsommeRm}) with $m$
standing for a couple of integers $m_\rho,m_\sigma$ with the same
parity. This substitution applies also to eq.(\ref{eqRmnew}), giving
\begin{eqnarray} \label{eqRmnewspin}
{\mathcal{R}}_{m_\rho,m_\sigma}& =&\int\-\-\-\int
\frac{V'(x)V'(y)}{(k_Fu)^2}e^{2im_\rho
\left[\phi_0(x)-\phi_0(y)\right]}\chi_{m_\rho,m_\sigma}(x,y).
\end{eqnarray}
The nonlocal function $\chi_{m_\rho,m_\sigma}(x,y)$ is now
more difficult to compute due to the different velocities of spin and
charge. But if we assume $u_\rho=u_\sigma$, it can be inferred from that
without spin (see appendix \ref{correlation}) by replacing
$$2m^2K\rightarrow m_\rho^2K_\rho+m_\sigma^2K_\sigma.$$ When the potential
has an extension less than $L$ and $L_{T,\nu}$ ($\nu=\rho,\sigma$), only the
local value is required as far as {\em the dominant term} is
concerned.\cite{ines_resonance} Then the leading term is similar to
eq.(\ref{Rlocal}) with $m$ replaced by the pair $(m_\rho,m_\sigma)$, thus
replacing $V_m$ by eq.(\ref{Vmspin}), and using
Eqs. (\ref{chihighspin},\ref{chilowspin}).

For $K_\rho<1/3$, we refer to figures \ref{fg:KlessT} and
\ref{fg:Kinterm}. Above (below) the crossover temperature $T_c(x)$, where
$x$ is the location of the barrier, the $4k_F$ (respectively $2k_F$)
backscattering gives the dominant contribution, and can be inferred from
Eqs. (\ref{chihighspin},\ref{chilowspin}) by taking
$(m_\rho,m_\sigma)=(1,1)$ (respectively $( 2,0)$) depending on whether
$T<T_L$ or $T>T_L$.
  
\subsection{Extended disorder: Conductance fluctuations}
For a random potential verifying eq.(\ref{gauss}), the leading term of
$\mathcal{R}$ is given explicitly in figure \ref{f:spindesordre}. Again the
case $K_\rho<1/3$ is more complicated: we have to distinguish three
temperature regimes, separated by $T_L$ than $T_c(0)$, eq.(\ref{Tc0}).

We can also compute the variance of $\mathcal{R}$, using
Eqs. (\ref{eqRmnewspin},\ref{gauss}). Similar steps to that
without spin can be carried on. While the powerlaw behavior
of $Var({\mathcal{R}})$ is different from that without spin,  the relative
fluctuations are the same, i.e. we have still eq.(\ref{ratiolow}) in the
low temperature limit, and eq.(\ref{ratiohighT}) (with a different function
$I$ that does not affect the order of magnitude) in the high temperature
limit.

Let us give the behavior of $Var({\mathcal{R}})$. In the low temperature
limit $T<T_L$, the inequality (\ref{encadrereflection}) holds again, showing
that $Var({\mathcal{R}})$ is of the same order as ${\mathcal{R}}^2$,
eq.(\ref{ratiolow}), given in fig.\ref{f:spindesordre}. In particular, in
the case $1/3<K_\rho<1$,
$$\sqrt{Var({\mathcal{R}})}\simeq \frac {u\tau_0}{l_e}\underline{T}_L^{2-K_\rho}.$$ For
$K_\rho<1/3$, this powerlaw holds at lower temperature $T<T_c(0)$,
eq.(\ref{Tc0}), but
$$\sqrt{Var{\mathcal{R}}}\simeq \frac
{u\tau_0}{l_e}\underline{T}^2\underline{T}_L^{4K_\rho-5}$$ for
$T_c(0)<T<T_L$.

In the high temperature limit $T>T_L$, restricting ourselves to $K_\rho<1$,
we can retain only the contribution of the bulk. Then we find
\begin{equation}
Var({\mathcal{R}})\sim\frac{u_\rho\tau_0
L}{l_e^2}\underline{T}^{m_\rho^2K_\rho+m_\sigma^2K_\sigma-3},
\end{equation}
up to a bounded function of both $k_FL_{T\nu}$ for $\nu=\rho,\sigma$. Again,
we have ignored the terms due to first and second derivatives of $\chi$, and
kept only the pair $(m_\rho,m_\sigma)$ that yields the dominant term. This
pair is equal to $(1,1)$ (respectively to $(2,0)$) for $K_\rho>1/3$
(respectively $K_\rho<1/3$) as is the case for the dominant contribution to
$\left\langle\mathcal{R}\right\rangle$. Remarkably, the ratio
(\ref{ratiohighT}) is {\em the same} as that without spin up to the function
$I$, and is $\ll 1$ in {\em any parameter range}.

\section{Discussion and Generalizations}
\label{secgeneral}
Different aspects of transport in an interacting wire with measuring leads
have been treated in this paper. In this section, we will compare the effect
of backscattering on the conductance to that obtained without leads, and then
discuss the results that can be generalized to smooth variations of
the interactions at the contacts.

In an infinite wire with repulsive interactions, as is thought to be
appropriate for a quantum wire, a unique local barrier is sufficient to
suppress transport completely.\cite{loiT,kane_furusaki} The finite length is
intuitively introduced as a cutoff, thus preventing the conductance to
vanish for weak enough backscattering potential or short enough
wire.\cite{apel_rice,kane_furusaki,ogata} With noninteracting leads and
spinless electrons, we recover the results of
refs.\onlinecite{ogata,kane_furusaki} over all the temperature range only for a
barrier whose separation from the contacts is of the order of the wire
length $L$, and under the additional condition $K_\rho>1/3$ for electrons
with spin.

If we consider extended disorder, the agreement holds for $K<(3+\sqrt
17)/2$ without spin, and for $1/3<K_\rho<1+\sqrt{2}$ with spin (see fig
\ref{f:spindesordre}). In this parameter range, the localization length we
infer from the breakdown of our perturbative evaluation of $\mathcal{R}$
coincides with that found in ref.\onlinecite{giamarchi_loc} in the weak
disorder limit.

We now discuss the extension of our results to smooth variations. As we
mentioned before, the discontinuous parameter model can be relevant for
edges states in the FQHE\cite{inhall,chamon_contacts}, but is not well
justified microscopically for quantum wires. Thus it is important to know
which results can be trusted for a more realistic interaction profile.

The detailed behavior of the transmission dynamics of an incident electron
was given previously for abrupt jumps.\cite{ines,ines_nato} By inspecting
the boson Green function, we can maintain the qualitative conclusion; in
particular, the transmitted charge and spin stay separated in the external
leads.

The correlation functions are directly related to this
process.\cite{these_annales} All the properties of $\chi$ found in
subsection \ref{properties} still hold.  To be specific, assume that the
leads are non interacting, $K_L=1$ and that $K$ decreases slowly enough from
$1$ to its bulk value $K$ inside the wire.  Again, for $K<1$,
[cf. Eqs. (\ref{chibulk}) and (\ref{majoration})] the tendency towards the
CDW holds inside the wire, but is reduced when one goes to the leads, until
it reaches its noninteracting value in the leads at a distance $u/(\omega
K)$ where $\omega$ is the maximum energy at hand, recalling the proximity
effect. Analogous conclusions hold for the superconducting fluctuations when
interactions are attractive, the coherence length being now $uK/\omega$.

In the presence of a non--random backscattering potential, the
renormalization equations (\ref{renormalisationspin}) are general. The role
of backscattering is determined by the density--density
correlations. Consider for instance a barrier at a point $x$. When $x$ is at
a distance of order $L$ of the contacts, the role of backscattering is the
same as that in ref.\onlinecite{kane_furusaki}. For $K<1$, $\mathcal{R}$
(see Eqs. (\ref{majoration}) and (\ref{Rlocal})) decreases when $x$ gets
closer to the contacts. The noninteracting leads (where a barrier is
marginal) tend to reduce the importance of the repulsive interactions when
one gets closer to them. Using the boson Green function properties on the
external leads (restricting ourselves to the case $K_\sigma=1$ with spin),
the temperature dependence of the correlation function corresponding to the
$2mk_F$ CDW at $T<T_L$ has a term $T^{m^2K_L-1}$. Thus for $K_L=1$, this
saturates for $m=1$ but goes to zero at zero temperature for $m\neq 1$. In
particular, the usual scaling argument inferring the low temperature
behavior $T<T_L$ from that at $T>T_L$ is justified only if the $2k_F$
contribution intervenes at all temperatures, which is not the case in our
model for spinful electrons with $K<1/3$, or for any parameter range when
the resonance in the presence of backscattering potential is achieved.

For a random potential, all the results can be generalized for repulsive
interactions. But the inequality (\ref{encadrereflection}) is more general
than that, and holds even if $K$ varies in any way inside the wire.

When spin is taken into account, and for $K_\sigma=1$, the tendency towards
a Wigner crystal is expected to be suppressed at low temperature, more
precisely the dominant tendency is that towards the $2k_F$ CDW at any
$T<T_c(0)$ [eq.(\ref{Tc0})]. This is because the dependence on temperature
of the $4k_F$ CDW is controlled by $K_L=1$, thus is in $T^{4K_L-1}=T^3$,
vanishing in the zero temperature limit.

 In the presence of Coulomb interactions, the ballistic conductance is still
independent on the interactions.\cite{ines_proc_arcs,these_annales,ines_bis}
One can make an analogy with the case $K_\rho<1/3$ to expect the transmitted
charge and spin to stay separated in the leads, and the tendency towards
Wigner Crystal to be suppressed by the leads at low enough temperature.

\section{Experiments}
We now comment on the experimental situation. In the high--temperature limit,
the ballistic value of the conductance of one-channel quantum wire $2e^2/h$
was observed by Tarucha {\it et al.},\cite{tarucha} as well as in the
remarkable clean wires of Yacoby {\it et al}.\cite{yacoby_second} But the
low temperature behavior is more tricky.

On the one hand, Tarucha {\em et al.} fitted the observed $\mathcal{R}$ with
the intuitive law $[T^2+T_L^2]^{(K_\rho -1)/2}$ of ref.\onlinecite{ogata},
and infer the parameter $K_\rho =0.7$. We however find a more complex
dependence of $\mathcal{R}$ on the temperature and the location of the
barriers. For instance, if the backscattering is merely due to the contacts,
the exponent $0.7$ would correspond to the local parameter $K_{a\rho
}=2K_\rho/(1+K_\rho)$, thus to $K_\rho \approx 0.5$. Of course we don't
expect the same combination of parameters to appear in the case of smooth
variations of $K$, but this serves as an illustration; as we discussed in
section \ref{secgeneral}, the effective parameter that controls the powerlaw
increases when one gets closer to the wire. Probably the potential
configuration depends on each wire, and more precise fit and evaluation of
$K_\rho $ are required. Even more caution is needed if the potential is
random. While the conductance is self--averaging at high temperature
(eq.(\ref{ratiohighT})), its fluctuations become more important at low
temperature, of the order of the reduction $\mathcal{R}$,
eq.(\ref{ratiolow}); then one has not to measure simply the absolute value
of $\mathcal{R}$, but also its fluctuations.

On the other hand, Yacoby {\em et al.}\cite{yacoby_second} fabricated high
quality quantum wires using a sophisticated technique. The most striking
result is that the conductance shows perfect plateaus as function of the
gate voltage. Their value is $G e^2/h$ per spin mode and channel where
$G\leq 1$. $G$ depends on the temperature and the wire length, but is
reproducible for different wires made in wells of the same width. This
indicates that the reduction in the conductance is not due to potential
fluctuations, but rather to the backscattering of electrons from the
two--dimensional gas when entering the wire. In the high temperature limit,
the one-channel wire conductance is $e^2/h$, ($G=1$) in accordance with
theoretical results. But $G$ decreases when one lowers temperature, or
increases the wire length, in a way similar qualitatively to the law
$T^{K-1}$ or $L^{1-K}$ for an interacting wire with parameter $K$.  The main
objection of Yacoby {\em et al.} against this TLL interpretation is that
temperature is scaled by the Fermi energy which varies with electron
density, in contradiction with the observed plateaus which are density
independent. Further, the parameter $K$ itself is expected to depend on the
density.

The importance of these two effects is however not so obvious to
assess. First, energy scales different from the Fermi energy might play the
role of the cutoff. Second the way $K$ depends on the density (i.e. the gate
voltage) is difficult to estimate: a fully microscopic calculation of $K$
would be necessary, including the capacitive effects of the gates. We have
however also to remark that our model is not the best suited to describe the
complicated geometry of the experiments. The transition between the
two--dimensional gas and the wire is not easy to describe, and is different
from the usual adiabatic transition. Thus we think model closer to the
actual experiment needs to be developed.\cite{remark_alekseev}

\section{Summary} 
In this paper we have studied the properties of an interacting wire
connected to measuring leads.  We have obtained a number of new results,
mainly concerning the suppression of the Wigner crystal and its effect on
the conductance with impurities, and the conductance fluctuations.  We have
also completed our study of the transmission process by taking spin into
account. We have shown here that the clean wire acts as a separator of the
charge and spin of an incident electron, which don't recombine in the
noninteracting leads. We can hope that experimental progress allows this
feature to be observable by appropriate spin and charge detectors. We have
also given a systematic computation of the nonlocal correlation functions
and their Fourier transform at temperature $T\gg \omega$.  This has allowed
to investigate both the dominant tendency in and outside the wire and to
study the role of impurities.

As first mentioned in ref.\onlinecite{ines}, we have explained in more
detail how the proximity effect shows up. For repulsive (attractive)
interactions, the charge density (pairing) fluctuations are enhanced in the
bulk of the wire, and decrease, while staying enhanced compared to the
noninteracting case, when one goes beyond the contacts.  But the novel
effect is that the leads influence the competition between different
tendencies in the wire at low enough temperature. This was shown to be the
case for very repulsive short range interactions $K<1/3$. Normally, a
tendency towards a Wigner crystal shows up, corresponding to the $4k_F$
CDW. When connected to leads, such tendency gets suppressed below a
space-dependent crossover temperature, and is replaced by the $2k_F$ CDW. In
particular, this allows for simultaneous $2k_F$ and $4k_F$ CDW in the wire
in a certain temperature range. $T_c$ is controlled by interactions, thus its
measurement could provide a way to measure $K$. The Wigner crystal shows up
more strongly for Coulomb interactions; we have not studied this case, but
we expect the same qualitative conclusions to hold.

We have also studied the role of a backscattering potential. We have
developed a formal framework to study its role for any non--translationally
invariant interactions, and explicitly at finite temperature. We have
derived the renormalization equations for a non--random potential $V$ with
any extension, but we expect interactions to be renormalized by a nonlocal
$V$. If $V(x)$ is weak enough for this effect to be neglected, the
conductance $g$ was expressed in detail in the discontinuous parameter case.
$g$ is controlled by the interactions, but is influenced by the external
leads, especially for very repulsive interactions because the leads suppress
the tendency towards the Wigner crystal below $T_c(x)$ [cf
Figs. \ref{fg:KlessT},\ref{fg:Kinterm}]. We have computed the conductance
fluctuations, whose behavior is nonuniversal.  But remarkably, the ratio
$Var({\mathcal{R}})/{\mathcal{R}}^2$ stays of the same order, whether the
spin is taken into account or not, or whether $K>1/3$ or not. It is $\sim
1/2$ for $T<T_L$, and $\sim L_T/L\ll 1$ for $T>T_L$. We can generalize the
main results to a smooth interaction profile near the contacts.

One of us (I. S.) would like to thank B. Dou\c cot, D. C. Glattli, T. Martin, D. L. Maslov and A. Yacoby for interesting discussions. 
 
\appendix

\section{Formal expression for the conductance}
\label{dyson}
We derive the analogue of a Dyson equation for the nonlocal conductivity. We
begin by spinless electrons, taking into account spin will the be
straightforward. We consider a general Hamiltonian
\begin{equation}\label{htot}
H_{tot}=H+ {\mathcal{V}}(\Phi),
\end{equation}
where $H$ is given by eq.(\ref{H}), and ${\mathcal{V}}(\Phi)$ depends
on $\Phi$, but not on its conjugate canonical momentum $
\Pi=\partial_x\Theta/\pi$.  Let us first give the general equation of motion
for $\Phi$.  Using $\left[ \Pi(x),\Phi(y)\right]=i\delta(x-y)$, any
functional $\mathcal{F}$ of $\Phi$ verifies
\begin{equation}\label{relbase}
\left[  \Pi,{\mathcal{F}}
\right] =-i\frac{\partial {\mathcal{F}}}{\partial \Phi}.
\end{equation}
This allows us to write the system
\begin{equation}
\left\{ 
\begin{array}{c}
\partial _t\Phi =\frac{\partial H_{tot}}{\partial \Pi }=\pi uK{ \Pi } \\
\partial _t\Pi =-\frac{\partial H}{\partial \Phi }=\frac 1\pi
\partial_x\left( \frac uK\partial _x\Phi \right) +{\cal I},
\end{array}
\right.  \label{piphi}
\end{equation}
where
\begin{equation}\label{calI}
{\cal I}(x)=\frac{\partial {\mathcal{V}}(\Phi)}{\partial \Phi (x)} 
\end{equation}
Note that $uK\mathcal{I}$ has the dimension of a force that intervenes in
the time derivative of $j=uK\Pi $. Eliminating ${\Pi }$ in
eq.(\ref{piphi}), the equation of motion for $\Phi $ reads
\begin{equation}
D_{xt}\Phi ={\cal I } \label{motI},
\end{equation}
where we denote
\begin{equation}\label{Dxt}
D_{xt}=\frac 1\pi \left[ -\partial _{tt}+uK\partial _x\left(
\frac uK\partial _x\right) \right].
\end{equation}
For an arbitrary nonlocal interaction potential $U(x,y)$, $D_{xt}$ contains
also an integral operator,\cite{ines_bis,these_annales} and all the
remaining analysis can be generalized to that case.

Note that without $\mathcal{V}$, we have ${\cal I } =0$, and thus recover
the equation of motion $D_{xt}\Phi=0$. In the following, the corresponding
correlation functions carry a subscript $0$, in order to distinguish them
from those computed with ${\cal V}$.

Next we derive a Dyson equation for the nonlocal conductivity ${\cal I}$,
eq.(\ref{calI}).  For any $A$ depending on $\Phi $ in a local or nonlocal
way, we can show, using eq.(\ref{relbase}) and eq.(\ref{motI}), that
\begin{equation}
D_{xt}\left\{ -i\theta(t)\left\langle \left[ \Phi (x,t),A\right]
\right\rangle \right\} =i\theta(t)\left\langle \left[ {\cal I}(x,t),A\right]
\right\rangle +\delta (t)\frac{\partial A}{\partial \Phi
(x)}\label{eqor}\end{equation} where $\theta$ is the Heaviside function.
The Fourier transform with respect to time of eq.(\ref{eqor}) yields
\begin{equation}
D_{x\omega }\left\langle \left\langle \Phi (x);A\right\rangle \right\rangle
_\omega =\left\langle \left\langle {\cal I}(x);A\right\rangle \right\rangle
_\omega +\frac{\partial A}{\partial \Phi (x)},  \label{eqw}
\end{equation}
where we adopt the notation \cite{gotze}
\begin{equation} \label{notationomega}
\langle \langle A(x);B(y)\rangle \rangle _{\omega} =-i\int_0^\infty e^{i\omega
t}\left\langle \left[ A(x,t),B(y,0)\right] \right\rangle_{H_{tot}}.
\end{equation}
We point out that the correlation functions are defined here with the total
Hamiltonian $H_{tot}$, eq.(\ref{htot}).  Similarly, we can show that
\[
D_{y\omega }\left\langle \left\langle A;\Phi (y)\right\rangle \right\rangle
_\omega =\left\langle \left\langle A;I(y)\right\rangle \right\rangle
_\omega +\frac{\partial A}{\partial \Phi (y)}. 
\]
Applying these two identities successively to the Green function $$
G(x,y,\omega )=\left\langle \left\langle \Phi (x);\Phi (y)\right\rangle
\right\rangle _\omega ,$$ we get
\[
D_{y\omega }D_{x\omega }G(x,y,\omega )=D_{y\omega }\delta
(x-y)+\frac{2i\omega}\pi f(x,y,\omega)
\]
where $f$ denotes
\begin{equation} \label{eq:fnonlocal}
f(x,y,\omega )=\frac \pi {2i\omega }\left[ \left\langle \left\langle {\cal I}
(x);{\cal I}(y)\right\rangle \right\rangle _\omega +\frac{\partial {\cal I}
(x)}{\partial \Phi (y)}\right] .
\end{equation}
In order to invert the relation above, and thus to obtain
$G_I(\overline{x},\overline{y},\omega)$ where $\overline{x},\overline{y}$
are arbitrary, we multiply its two members by $G_0(\overline{y},y,\omega)$,
integrate over $y$, and use $$D_{y\omega
}G_0(\overline{y},y,\omega)=\delta(y-\overline{y}).$$ Then we multiply both
sides by $G(\overline{x},x,\omega )$ and integrate over $x$. We get
\begin{eqnarray}\label{dysonG}
G(\overline{x},\overline{y},\omega )&=&G_0(\overline{x},\overline{y} ,\omega
)+\frac{2i\omega}\pi\int\-\int dx dy \nonumber\\&&G_0(\overline{x},x,\omega
)G_0(\overline{y},y,\omega )f(x,y,\omega ).
\end{eqnarray}
We can also express this equation in terms of the dimensionless nonlocal
conductivity (measured in units of $e^2/h$), eq.(\ref{sigmaG}): 
\begin{eqnarray}\label{cond}
\sigma(\overline{x},\overline{y},\omega )&=&\sigma_0 (\overline{x},
\overline{y},\omega )-\int\int dxdy\\\nonumber&&\sigma_0
(\overline{x},x,\omega )\sigma (\overline{y},y,\omega)f(x,y,\omega ).
\end{eqnarray}
The conductance measured in units of $e^2/h$ is given by eq.(\ref{gsigma}),
\begin{equation}g=\lim_{\omega \rightarrow 0}\sigma
(\overline{x},\overline{y},\omega )=\sigma (\overline{x},
\overline{y})\end{equation} independent of its arguments, which ensures
current conservation. It is also real because it is a correlation function
of two Hermitian fields. Indeed, for two hermitian operators $A,B$, we have
$$\left( \left\langle \left\langle A;B\right\rangle \right\rangle _0\right)
^{\dagger}=i\int_0^\infty \left\langle \left[ B^{\dagger},A^{\dagger}\right]
\right\rangle =\left\langle \left\langle A;B\right\rangle \right\rangle _0
\;\;\;\mbox{real}.$$ Besides,
\[
\frac d{d\omega }\left\langle \left\langle A;B\right\rangle \right\rangle
_{\omega =0}=\int_0^\infty t\left\langle \left[ A,B\right] \right\rangle dt
\;\;\;\mbox{is purely imaginary}
\]
Thus $\lim_{\omega \rightarrow 0}\left\langle \left\langle {\cal I}(x);
{\cal I}(y)\right\rangle \right\rangle _\omega $ is real while its
derivative at zero frequency is purely imaginary because ${\cal I}$ is
hermitian.  If one supposes $\sigma {}(\overline{x},\overline{y})$ finite,
one has
\[
\int dx\int dy\left[ \left\langle \left\langle {\cal I}(x);{\cal I}
(y)\right\rangle \right\rangle _0+\frac{\partial {\cal I}(x)}{\partial \Phi
(y)}\right] =0.
\]
Then the zero frequency limit of eq.(\ref{cond}) yields the exact
conductance measured in units of $e^2/h$
\[
g=g_0\left[ 1- g_0\int dx\int dyf(x,y)\right] .
\]
where
\[
f(x,y)=\frac \pi 2Im\frac d{d\omega }\left\langle \left\langle {\cal I}(x);
{\cal I}(y)\right\rangle \right\rangle _{\omega =0} 
\]
is the zero--frequency limit of eq.(\ref{eq:fnonlocal}) and $g_0$ is the
dimensionless conductance in the absence of $\mathcal{V}$.  If one takes
spin into account, ${\mathcal{V}}$ can depend on the spin field
$\Phi_\sigma$ and its conjugate canonical momentum $\Pi_\sigma$, both
commuting with the field $\Phi_\rho$.  This is the case for instance for
backscattering potential, where the contribution of the $(m_\rho,m_\sigma)$
harmonics of the density (\ref{rhospin}) to $\mathcal{I}$ reads, after
integration by parts: $\int
V'(x)\exp\left({2im_\rho\widetilde{\Phi}_\rho+2im_\sigma\Phi_\sigma}\right)$.

\section{Correlation functions}
\label{correlation}
We will show how to compute the imaginary part of the Fourier transform of
the nonlocal correlation function (\ref{Xm}) at frequency $\omega
\ll T$, i.e. $\chi(x,y)$ [eq.(\ref{imaginary})].

First we clarify the way analytic continuations from Matsubara to real
frequencies are done when we regularize by imposing the imaginary time to
obey $\tau_0\leq \tau\leq \beta-\tau_0$.  Consider any general operators $A$
and $B$. Using the equivalent of the notation (\ref{notationomega}) for
the Matsubara frequency $i\omega_n$, we have
\begin{eqnarray}\label{AB} 
\langle \langle A;B\rangle \rangle _{i\omega _n} &=&\int_{\tau _0}^{\beta
-\tau _0}e^{i\omega _n\tau }\left\langle A(\tau )B(0)\right\rangle
\nonumber\\ &=&\int_0^\infty d(it)\left\{ -e^{i\omega _n(it+\tau
_0)}\left\langle A(it+\tau _0)B(0)\right\rangle
\right.\nonumber\\&&\left. +e^{i\omega _n(it+\beta -\tau _0)}\left\langle
A(it+\beta -\tau _0)B(0)\right\rangle \right\} \nonumber \\ &=&
-i\int_0^\infty dte^{-\omega _nt}\left\{ e^{i\omega _n\tau _0}\left\langle
A(it+\tau _0)B(0)\right\rangle\right.\\ \nonumber&&\left. -e^{-i\omega
_n\tau _0}\left\langle B(-it+\tau _0)A(0)\right\rangle \right\},\nonumber
\end{eqnarray}
where we have deformed the integration contour in the complex plane, and
used
\[
\left\langle A(\beta -z)B(0)\right\rangle =\left\langle
B(z)A(0)\right\rangle 
\]
Let us now specialize to the case $$A=B^{\dagger}=e^{2im\Phi}$$ where $m$ is
 normally an integer but can be any real, and consider $X_m$, eq.(\ref{Xm})
 \begin{equation} X_m(x,y,\tau )=\left\langle T_\tau
 A^{\phantom{\dagger}}(x,\tau )A^{\dagger}(y,0)\right\rangle.
\end{equation}
Then we have
\begin{equation}
\left\langle T_\tau A(x,\tau
)A^{\dagger}(y,0)\right\rangle =\left\langle T_\tau A^{\dagger}(x,\tau
)A(y,0)\right\rangle .
\end{equation}
This is because the transformation: $ \Phi\to-\Phi $ leaves $H$ invariant,
but $A\to A^{\dagger}$. Besides, $X_m$ is invariant when
exchanging $x$ and $y$ because of time reversal symmetry .

Equation (\ref {AB}) becomes, after the analytic continuation
$i\omega_n\rightarrow \omega+i\delta$,
\begin{eqnarray}
X_m(x,y,\omega )&=&-i\int_0^\infty dte^{i\omega t}\left\{ e^{\omega \tau
_0}\left\langle A(x,it+\tau
_0)A^{\dagger}(y,0)\right\rangle\right.\nonumber\\
&&\left.-e^{-\omega \tau
_0}\left\langle A^{\phantom{\dagger}}(x,-it+\tau
_0)A^{\dagger}(y,0)\right\rangle \right\}.\label{xxxx}
\end{eqnarray}
The derivative with respect to $\omega$ yields, after changing $t\to -t$ in
the second term above,
\begin{equation}
\left. \frac{dX_m}{d\omega }\right|_{\omega =0}=-i\int_{-\infty }^\infty
(it+\tau _0)X_m(x,y,it+\tau _0),  \label{eq:dRO}
\end{equation}
where $X_m(x,y,z)=\left\langle
A(x,z)A^{\dagger}(y,0)\right\rangle $.  According to
eq.(\ref{eq:dRO}), and after the change of variables $t\rightarrow \pi Tt$,
\begin{equation}
\left. -\frac{d\chi _m}{d\omega }\right|_{\omega =0}=\frac{1}{(\pi
T)^2}\int_{-i\infty +\underline{T} }^{i\infty +\underline{T}}z\exp \left[
-2m^2U(x,y,z)\right] dz , \label{dchim}
\end{equation}
where $\underline{T}=\pi T\tau_0$, and $U$ is given by eq.(\ref{Utau}). 

Owing to the analytic properties of $U$, we can move the integration contour
in eq.(\ref{dchim}) as explained now. Let us write the integral in
eq.(\ref{dchim}) as $$\int_{-i\infty +\underline{T} }^{i\infty
+\underline{T}}zX_m(z),$$ where the spatial arguments are implicit. $U$ is
analytic on the complex band $\underline{T}\leq {\rm Re}(z)\leq \pi /2$, as well as
$zX_m(z)$.

Let $z=\mu +i\nu$. We then have $\underline{T}\leq \sin \mu \leq 1$.  We can
show that for $\nu \rightarrow \infty $, $$X_m(z)\sim e^{-m^2K_L\vert
\nu\vert}.$$
 Then we can translate the contour of integration by $\pi /2-\underline{T},$
thus making it along $z=it+\pi/2 $. On the other hand, we can show that
$X_m(it+\pi/2)$ is an even function of $t$, thus $$\int_{-\infty }^{+\infty
}t X_m(it+\pi/2)=0.$$ Finally, the integral (\ref{dchim}) becomes
\[
-\left. \frac{dX _m}{d\omega }\right|_{\omega =0}=\frac{a} {2\pi
T^2}\int_{-i\infty +\pi /2}^{i\infty +\pi /2}dze^{-2m^2U(z)}
\]
This form is much more convenient than (\ref{dchim}) for three reasons: the
integrand is real, the singularities are avoided, and approximations are
easier to make.  Thus the dimensionless function $\chi$,
eq.(\ref{imaginary}), corresponding to $m=1$ can be expressed as
\begin{equation}\label{dchi}
\chi\left( x,y\right)=\frac{1}{2\underline{T}^{2}}\int_{-i\infty +\pi
/2}^{i\infty +\pi /2}X (x,y,\tau)d\tau.
\end{equation}
This gives eq.(\ref{chic}):
\begin{equation}
\chi\left( x,y\right)=\frac{1}{\underline{T}^2 }C(x,y)e^{-2G(x,x,\tau_0
)}e^{ -2G(y,y,\tau_0 )}, \label{apchic}
\end{equation}
where 
\begin{equation}
C(x,y)=\int_0^\infty d\zeta \exp \left[4G(x,y,i\zeta +\pi /2)\right].
\label{Gint}
\end{equation}
 All the $\chi_m$ can be obtained by multiplying $G$ by $m^2$.

\subsection{Discontinuous parameters}
Let us give the function $G$ in the case of discontinuous interaction
parameters\cite{these_annales}
\begin{eqnarray}
\nonumber
\lefteqn{-2G(x,y,z)/K =h\left( z,v_{-}\right) +\gamma h\left( z,v_{+}\right)
 +\sum_{r=\pm 1}\sum_{p=1}^\infty} \\ 
\label{apGz}
&&\left[ \gamma ^{2p}h\left(
 z,2pv_L+rv_{-}\right) +\gamma ^{2p+r}h\left( z,2pv_L+rv_{+}\right)
 \right]
\end{eqnarray}
with 
\begin{equation} \label{apdefh}
h(z,v)=\frac 12\log \left[ \sin ^2\pi T z+\sinh^2v\right] ,
\end{equation}
and 
\begin{eqnarray}
v_{-} &=&\pi T\left| x-y\right| /u  \nonumber \\
v_{+} &=&\pi T\left( L-\left| x+y\right| \right) /u  \nonumber \\
v_L &=&\pi T L/u.  \label{lesv}
\end{eqnarray}
$\gamma$ is given by eq.(\ref{gamma}). It is worth noting that one can
deduce $G$ from $\sigma(x,y,\omega)$ as explained in
ref.\onlinecite{these_annales}, and that $\sigma(x,y,\omega)$ is related to
the transmission process.\cite{ines}

\subsubsection{High temperature limit}
Let us first consider the limit $T\gg T_L$. We can show in detail that a
good approximation of the integral $C(x,y)$, eq.(\ref{Gint}), can be
obtained if we approximate $G$ by
\begin{equation}
-2G(x,y,z)\simeq K\left\{ h\left( z,v_{-}\right) +\gamma \left[ h\left(
z,v_{+}\right) -v_{+}+\log2\right]\right\}.  \label{Gzfaible}
\end{equation}
There is no coherence between the two contacts separated by $L\gg L_T$, so
that only the role of one reflection matters. The integral can be expressed
as
\begin{eqnarray}
C\left( x,y\right)& =&\left( 1+e^{-2v_{+}}\right) ^{-2\gamma K}\frac{B\left(
\frac 12,K_a\right) }{2\cosh ^{2K}v_{-}}\nonumber\\ &&F_1\left( \frac
12,K,\gamma K,\frac 12+K_a;\tanh ^2v_{-},\tanh ^2v_{+}\right),\nonumber\\&&
\label{CF1}
\end{eqnarray}
where $F_1$ is the hypergeometric function generalized to two
variables,\cite{gr} and $K_a$ given by eq.(\ref{Ka}).

 In the limit of an infinite wire with constant parameters, we can set
$K=K_L$, i.e. $\gamma=0$. Then $C(x,y)$ is a function of $x-y$, or of $v$
[eq.(\ref{lesv})] and becomes
\begin{equation}
C_K\left( v\right)=B( \frac 12,K) {_2F_1}\left( K,K,\frac 12+K;-\sinh
^2v\right), \label{CKv}
\end{equation}
where $B$ is the Beta function and $_2F_1$ is the hypergeometric function of
one variable.

At distances far from the contacts compared to $L_T$, thus for $v_+\gg 1$,
we have
$$C(x,y)\simeq C_K(v_-).$$  
In order to express $\chi$, eq.(\ref{apchic}), we need also to exponentiate
eq.(\ref{Gzfaible}) for $x=y$, and $z=\tau_0$. Thus one gets $\chi(x,y)$ for
any $x,y$.

\paragraph{Local value} 
For $\vert x-y\vert\ll L_T$, we can check that $C(x,y)\simeq
C(x,x)$. Nevertheless, for $e^{-2G(x,x,\tau_0)}\simeq e^{-2G(y,y,\tau_0)}$
to hold, we must have $t_x,t_y\gg1$. Thus strictly speaking the property
(\ref{chiloc}) (now $L_{min}=L_T$ since we are in the high temperature
limit) holds only for $t_x\gg1,t_y\gg1$. But this is related to the abrupt
jump of parameters at the contacts, thus we don't expect it to hold for
smooth variations, where eq.(\ref{chiloc}) can be generalized. Besides
bosonization describes large separations compared to $u\tau_0$, thus the
variations of $K$ as well as the behavior for $t_x\ll 1 $ are not relevant.

Recall that $t_x$ is the time it takes for a ``quasiparticle'' to go from $x$
to the closest contact, measured in units of the cutoff $\tau_0$
[eq.(\ref{tx0})]:
$$t_x=\frac{a-|x|}{u\tau_0}.$$

Let us write the local value of $\chi$:
\begin{eqnarray}\label{Upsilonxxtext}
\chi\left( x,x\right)&=&\frac 12\underline{T} ^{2\left( K-1\right) }B\left(
\frac12,K_a\right)\\\nonumber&& \left[ \underline{T} ^2+\left(
1-\underline{T} ^2\right) \tanh^2( \pi \underline{T}t_x)\right]
^{K_a-K}\\\nonumber&&{_2F_1}\left( \frac 12,K_a-K,K_a+\frac 12,\tanh ^2(\pi
\underline{T}t_x)\right).
\end{eqnarray}
 The Hypergeometric function $_2F_1$ is unity at the contact ($v=0$) and
 stays bounded and nonvanishing all over the wire. At $x=a$,
 \begin{equation} \chi(a,a)=\frac 12B\left( \frac 12,K_a\right)
 \widetilde{T} ^{2(K_a-1)}.\label{chia}
\end{equation}
At $x=0$, an analogous expression holds with $K_a\rightarrow
K$. Besides, we can show that:\cite{these_annales}

$$\chi(a,a)\leq \chi(x,x)\leq \chi(0,0)$$
for $K<K_L$, while the inverse holds for $K>K_L$.

We note finally that the simplified expression (\ref{Vrenorm}) was obtained
by ignoring the hypergeometric function in eq.(\ref{Upsilonxxtext}), as well
as constant factors, and taking $t_x\gg1$. But we can check that
$\chi(a,a)$, eq.(\ref{chia}), can be recovered from eq.(\ref{Vrenorm}) by
letting $t_x=1$.

\paragraph{Expression of $\chi(a,x)$}

 Another interesting expression is that of $\chi(a,x)$ for $1\ll t_x\ll
 \underline{T}^{-1}$, thus $v_-\ll 1$ in eq.(\ref{CF1}):

\begin{equation} \label{chiayasymptotique}
\chi\left( a,x\right)\simeq \frac 12B\left( \frac12,K_a\right) \underline{T}
^{2(K_a-1)}{t_x}^{K_a-K}
\end{equation}

\subsubsection{Low temperature limit}
The computation of $C(x,y)$ is now much easier.  Since in eq.(\ref{Gint})
$z=i\zeta+\pi /2$, $-\sin z=\cosh \zeta\geq 1\gg v_L\simeq \sinh v_L$. Then
one can approximate, in eq.(\ref{apGz}),
\begin{equation}
2G(x,y,i\zeta+\pi /2)\simeq -2K_L\log \left[ \cosh t\right] , \label{G}
\end{equation}
thus eq.(\ref{Gint}) reduces simply to
\begin{equation}
C(x,y)=\int_{-\infty }^\infty \frac{dt}{\cosh^{2K_L}t}=\frac 12B\left(
\frac 12,K_L\right),  \label{Clow}
\end{equation}
independent of $x,y$. $C(x,y)$ is the same as $C(x,x)$ evaluated in an
infinite wire with parameter $K_L$. But $\chi$ [eq.(\ref{apchic})] has also
parts depending on $x$ and $y$ due to $G$. By letting $x=y$ in
eq.(\ref{apGz}) we get
\begin{equation}
\underline{T}^{-1}e^{-2G(x,x,\tau_0)}\simeq \underline{T}
^{K_L-1}\underline{T}_L^{ K_a-K_L }\left( 1+t_x^2 \right)
^{(K_a-K)/2},\label{Glow}
\end{equation}
up to a bounded and slowly varying function of $t_x$. Inserting this
expression for each of $x,y$ as well as eq.(\ref{Clow}) in
eq.(\ref{apchic}), we obtain $\chi(x,y)$, factorizing as in
eq.(\ref{decouple}) because $L\ll L_T$.

Note that taking $t_a=0$ in eq.(\ref{Glow}) or ignoring the $1$ in front of
$t_x^2$ then taking $t_x=1$ gives the same result for $\chi(a,a)$. The
behavior for $t_x\ll 1$ is not very relevant since it depends closely on the
unrealistic abrupt profile of $u,K$, and also because the TLL model is valid
at distances much larger than $u\tau_0$, in particular $t_x\gg 1$.

\section{The suppression of the Wigner crystal}
\label{appwigner}
Here we discuss in detail the competition between the $2k_F$ CDW and the
$4k_F$ CDW, separating the high-- and low--temperature regimes. For
simplicity, we restrict ourselves to the case $K_\sigma=1$, and we drop the
index $\rho$ from all the parameters or length scales. Note that the
effective local parameter at the contacts, eq.(\ref{Ka}), for the spin
degrees of freedom is also equal to one: the correlation function
$U_\sigma(x,y,\tau)$ is now equal to that in a noninteracting wire for any
$x$ and $y$.

\subsubsection{High temperature limit}
We consider here $T\gg \max (T_{L,\rho},T_{L,\sigma})$. Recall that the
correlation functions at distances far from the contacts compared to
$(L_{T\rho},L_{T\sigma})$ are identical to those in an infinite TLL,
eq.(\ref{chispinbulk}). At the contacts, the equivalent TLL has a parameter
$K_{a}$, eq.(\ref{Ka}).

For any $x$, let us use eqs.(\ref{factorspin},\ref{integration}) to obtain
the ratio between the $2k_F$ and $4k_F$ CDW correlation functions,
\begin{equation}\label{ratio_correlation_high}
\lambda(x)=\frac{\chi_{1,1}}{\chi_{2,0}}=\underline{T}^{1-3K}\left(\tanh
\underline{T}t_{x,\rho}\right)^{-3\gamma K},
\end{equation}
where $\gamma$ is given by eq.(\ref{gammanu}), and
$t_{x}=(a-|x|)/u_\rho\tau_0$ is the time it takes for a charge excitation
emanating at $x$ to reach the closest contact. The $4k_F$ CDW dominates the
$2k_F$ one wherever $\lambda(x)\ll 1$. Thus:
\begin{itemize}
\item
 For $Tt_{x}\gg1$, the $4k_F$ CDW is dominant for any $K<1/3$ as in an
 infinite TLL, because $\tanh A\simeq 1$ for $A\gg 1$.
\item
 For $t_x\sim 1$, i.e. at the contacts, $\lambda(a)\underline{T}^{1-3K_a}$:
 the $4k_F$ dominate for $K_a<1/3$, i.e. for $K<1/5$.
\item
 For any $t_{x} \underline{T}\ll 1$, we can replace $\tanh\epsilon \simeq
\epsilon$ if $\epsilon\ll 1$, and the ratio (\ref{ratio_correlation_high})
becomes
\begin{equation}\label{ratio_small_tx}
\lambda=\underline{T}^{1-3K_a}t_{x}^{-3\gamma K}.
\end{equation}This expression holds for $t_{x}\gg1$, but the
computation can be carried out for any $x$ (see appendix \ref{correlation}).

Since $K<1/3<1$, $\gamma>0$, $t_x^{-3\gamma K}$ increases when one gets
closer to the contacts, being $\ll 1$ for $t_x\gg1$, and of order $1$ at the
contacts. This means that the importance of the $2k_F$ compared to the
$4k_F$ is enhanced as one gets closer to the leads.  We have to discuss
whether $K<1/5$ or $K>1/5$. In the former case, $\lambda(x)\ll 1$ (cf
eq.(\ref{ratio_small_tx})). In the latter case $\lambda$ reaches unity for
\begin{equation}\label{xch}
t_{c}(T)=\underline{T}^{\frac{1-3K_a}{3\gamma K}}.
\end{equation}
Thus, for $t_x<t_{c}(T)$, the $2k_F$ CDW dominates (see
fig.\ref{fg:Kinterm}).
\end{itemize}
To summarize the high temperature regime, the $4k_F$ ($2k_F$) CDW is dominant
all over the wire for $K<1/5$ ($K>1/3$). In the intermediate
parameter region, i.e. for $1/5<K<1/3$, the $4k_F$ CDW dominates only
for points far enough from the contacts, i.e. for $t_x>t_c(T)$, eq.(\ref{xch}).
\subsubsection{Low temperature limit}
Consider now the low temperature limit, $T\ll
\inf(T_{L,\rho},T_{L,\sigma})$. Let us write again eq.(\ref{chilowspin})
\begin{equation}\label{lowchispin}
\chi_{m_\rho,m_\sigma}(x,x)\simeq \underline{T}^{m_\rho^2+m_\sigma^2-2}
\underline{T}_{L,\rho}^{-m_\rho^2\gamma}t_{x}^{m_\rho^2\gamma K}.
\end{equation}
This holds for $t_{x,\rho}\gg1$, and the value at the contact can be
recovered by simple substitution $t_{x}=1$ in eq.(\ref{lowchispin}) as we
can check from explicit computation for any $x$.\cite{these_annales} The
essential difference between $\chi_{1,1}$ and $\chi_{2,0}$, due to the
leads, is through the temperature dependent term $T^2$ in $\chi_{2,0}$ that
vanishes in the zero-temperature limit, while $\chi_{1,1}$ is
temperature-independent. Then the $2k_F$ CDW dominates the $4k_F$ CDW in the
zero-temperature limit all over the wire for any parameter $K$.  For
$K>1/3$, the $2k_F$ dominates everywhere at any temperature.  For $K<1/3$,
the $4k_F$ CDW dominates in the high temperature limit. Thus there is a
crossover temperature we determine now.

Let us write the ratio of the corresponding correlation functions at finite
temperature, using eq.(\ref{lowchispin}),
\begin{equation}\label{ratio_correlation_low}
\lambda(x)=\frac{\chi_{1,1}}{\chi_{2,0}}=
\underline{T}^{-2}t_{x}^{-3(1-K)}\left(\underline{T}_{L}t_{x}\right)^{3\gamma}.
\end{equation}
\label{transition}
Again, we have to distinguish the cases $K<1/5$ and $K>1/5$:
\begin{itemize}
\item
For $K<1/5$, we can easily get the crossover temperature by letting
$\lambda(x)\sim1$,
\begin{equation}\label{Tcl}
\underline{T}_c(x)=\left(\frac{\underline{T}_L}{t_x^K}\right)^{\frac 3
2\gamma}.
\end{equation}
We can check that for any $x\in [-a,a]$, $T_c\leq T_L$, and $T_c$ is a
decreasing function of $t_x$ (fig.\ref{fg:KlessT}).
\item
For $1/5<K<1/3$, the same expression for the crossover temperature
holds, eq.(\ref{Tcl}). But it is inside the actual range of temperatures,
i.e. $T_c<T_L$ only for points less than $x_0$ where $x_0$ verifies
$T_c(x_0)=T_L$, i.e. given by
\begin{equation}\label{tx0}
t_{x0}={\underline{T}}_L^{\frac{1-3K_a}{3\gamma K}}.
\end{equation}
\end{itemize}
Let us now analyze the high and low temperature results together.
\begin{itemize}
\item
For $K<1/5$, the $4k_F$ CDW dominates for $T>T_c(x)$, eq.(\ref{Tcl}), while
the $2k_F$ CDW dominates below (fig.\ref{fg:KlessT}). In particular, for a
given temperature in the range $[T_c(0),T_c(a)]$, there is a spatial
crossover between the two tendencies occurring at $x(T)$, one can obtain by
inverting eq.(\ref{Tcl}).
\item
For $1/5<K<1/3$, we can invert the relation (\ref{xch}) to get the crossover
temperature higher than $T_L$ for $\vert x\vert >x_0$, and we can (which is
reassuring) check that the point where it reaches $T_L$ is again given by
$x_0$, eq.(\ref{tx0}), showing that we have a continuous function $T_c(x)$
given by
\begin{eqnarray}\label{TcKmore}
T_c(x)&=&T_L\left(\frac{t_x}{t_{x_0}}\right)^{-\frac 3 2\gamma K}
{\mbox\,\,\, for \,\,\,} t_x>t_{x_0}\\\nonumber
&=&T_L\left(\frac{t_x}{t_{x_0}}\right)^{-\frac{3\gamma K}{3K_a-1}}
{\mbox\,\,\, for \,\,\,} t_x<t_{x_0}.
\end{eqnarray}
Note that the $2k_F$ CDW dominates for any temperature at the contacts:
$t_x\sim1$, thus $T_c(a)$ in the second line diverges
(fig.\ref{fg:Kinterm}).
\end{itemize}

\begin{figure}[htb]
\epsfig{file=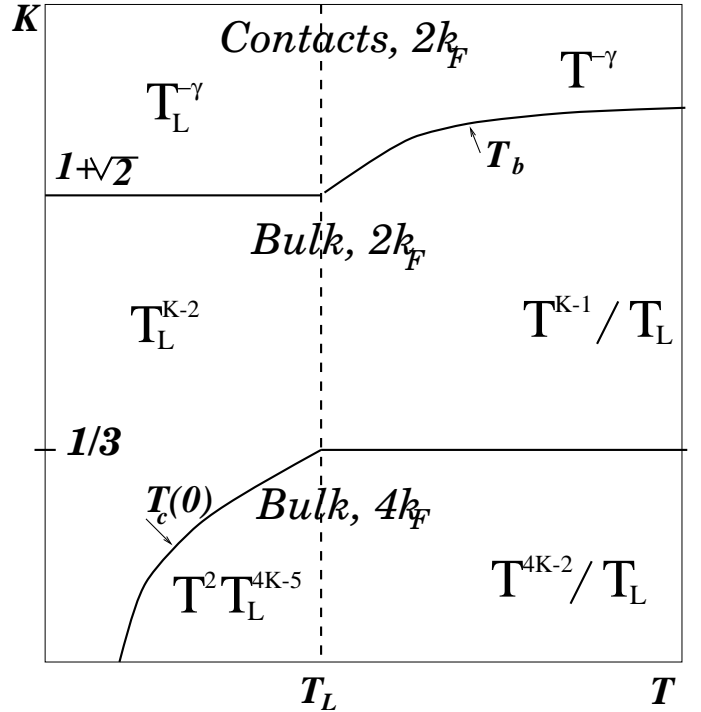,width=9cm}
\caption{ ${\mathcal{R}}=1-g/(2e^2/h)$ multiplied by $l_e/u\tau_0$ for
extended disorder as function of the temperature (the x-axis) and $K_\rho$
(the subscript $\rho$ is dropped for simplicity). In each of the three
regions, we indicate whether the dominant contribution comes from the bulk
or the contacts, and from the $2k_F$ or the $4k_F$ backscattering. The
impurities in the bulk (controlled by $K_\rho$) dominate those closer to the
contacts (controlled by $K_{a\rho}$) either at any $T$ if $K_\rho<1+\sqrt 2$
or at $T>T_b=T_L^{-1/(\gamma_\rho K_\rho)}$ if $K_\rho >1+\sqrt{2}$. For
$K_\rho<1/3$, the $4k_F$ backscattering dominates only at high temperature
$T>T_c(0)$, eq.(\ref{Tc0}).}
\label{f:spindesordre}
\end{figure}

%\bibliography{../../revues,../../data,annexe}

\begin{thebibliography}{10}
\bibitem{gremark}
We subtract a formally infinite constant from $G$. As all results depend only
  on differences of $G$'s, this is of no importance.
\bibitem{maslov_proximity}
In these previous works, we have studied the pairing correlation functions. See
  also a recent paper by D.~L. Maslov and P. Goldbart, cond-mat/9711070, where
  the local density of states is computed.
\bibitem{remarkkf}
We have to precise that our discussion is different from that of
  Ref.\onlinecite{ines_resonance} where the competition between $2k_{F\text{
  }}$ and $4k_F$ is investigated for electrons without spin, in a wire without
  leads.
\bibitem{remark_m}
Strictly speaking, this is true only for $\vert m_\rho\vert= \vert
  m_\sigma\vert$. Otherwise, the function $V'(x)$ has to be replaced by a
  higher order term that emanates from the integration of
  eq.(\ref{higher_m_spin}).
\end{thebibliography}

\end{document}